\documentclass[final,1p,times,twocolumn]{elsarticle}
\usepackage{amssymb}
\usepackage{caption}
\usepackage{amsmath}
\usepackage{varwidth}
\usepackage{ifthen}
\usepackage{tabularx} 

\graphicspath{{./figures/}{./figures/eps/}{./figures/pdf/}}
\usepackage[colorlinks,
            citecolor=blue,
            linkcolor=blue,
            urlcolor=blue,
            pdfstartview=FitH,
            ]{hyperref}
\usepackage{comment} 
\usepackage{soul,xcolor} 
\usepackage{upgreek} 
\usepackage{bm} 
\usepackage{soul} 
\usepackage{svg} 
\usepackage[version=3]{mhchem} 
\newcommand\Niaa{\ce{^{58}Ni}($\alpha$,$\alpha^\prime$)\ce{^{58}Ni}}

\journal{Nuclear Instruments and Methods in Physics Research A}
\begin{document}

\sloppy
\setstcolor{red} 


\begin{frontmatter}

\title{Applying Gaussian Mixture Models to Track Reconstruction in Inelastic Scattering Experiments with Active Targets\tnoteref{label1}}

\title{}

\author[KUL]{A.~Arokiaraj\corref{cor1}}
\author[CERD]{M.~Latif}
\author[KUL]{R.~Raabe}
\author[CEA]{D.~Thisse}
\author[CEA]{M.~Vandebrouck}

\affiliation[KUL]{%
  organization={KU Leuven, Institute for Nuclear and Radiation Physics},
  postcode={3001},
  city={Leuven},
  country={Belgium}
}
\affiliation[CERD]{%
  organization={Division of Applied Nuclear Science and Technology, Centre for Energy Research and Development, Obafemi Awolowo University},
  postcode={220005},
  city={Ile-Ife},
  country={Nigeria}
}
\affiliation[CEA]{%
  organization={Irfu, CEA, Université Paris-Saclay},
  postcode={91191},
  city={Gif-sur-Yvette},
  country={France}
}

\cortext[cor1]{Corresponding author}

\begin{abstract}
Active targets such as ACTAR TPC~\cite{mauss2019commissioning} are well suited for studying giant resonances in unstable nuclei via inelastic scattering in inverse kinematics.
A key challenge in such measurements is the detection of low-energy ejectiles emitted at small angles relative to the beam direction.
Accurate reconstruction of these tracks is essential for disentangling different resonance modes.
Probabilistic models such as the Gaussian Mixture Model (GMM) are particularly effective in capturing the complex covariance structures characteristic of the beam-recoil interface in narrow-angle events.
In this work, we present a track reconstruction approach based on the GMM, specifically designed for inelastic scattering experiments with active targets.
Special emphasis is placed on the treatment of low-energy tracks.
The proposed method is demonstrated on simulated data of the \Niaa\ reaction at an incident energy of $E=49$~MeV/nucleon, generated under conditions representative of the experiment carried out at GANIL for the same reaction.
\end{abstract}


\end{frontmatter}


\section{Introduction}
\label{intro}
The experimental investigation of giant resonances using active targets has gained significant momentum over the past two decades.
One of the principal objectives of such studies is the excitation of the isoscalar giant monopole resonance, due to its strong correlation with the incompressibility of nuclear matter, $K_\infty$~\cite{blaizot1980nuclear,harakeh2001giant}.
These measurements are progressively being extended to nuclei far from stability.
The resulting data provide valuable insight into the role of neutron--proton asymmetry in the compressibility parameter and help constrain uncertainties in the incompressibility of infinite nuclear matter~\cite{garg2018compression}.
In parallel, theoretical predictions have suggested the appearance of soft monopole resonances in neutron-rich systems, such as the nickel isotopic chain~\cite{khan2011low,gambacurta2019soft}.

Experiments are typically performed using inelastic scattering at energies of a few tens of MeV/nucleon.
For investigations involving unstable nuclei, the arrangement is in inverse kinematics, where the heavier nucleus serves as the projectile and impinges on protons, deuterons, or $\alpha$ particles.
These measurements rely on the detection of the light recoil.

The cross section is largest at small center-of-mass angles, for which the light recoils are emitted at small forward angles and with kinetic energies ranging from 10~keV to 5~MeV, well below the threshold required to exit a solid target.
Gaseous targets are therefore employed; when the gas also serves as a detection medium, allowing the trajectories of charged particles to be reconstructed from their ionization footprint in a time-projection chamber (TPC), such systems are referred to as ``active targets''~\cite{Bazin2020}.

Active-target TPCs offer nearly $4\pi$ angular coverage and enable three-dimensional reconstruction of ion tracks.
These detectors employ an electric field to drift ionization electrons to a segmented readout plane, where amplification and signal collection occur.
The two-dimensional projection of the tracks is obtained from the segmented plane, while the third dimension is extracted from the electron drift time.

Extracting reaction kinematics from particle tracks is a challenging problem, which for low-energy nuclear reactions such as those considered here goes beyond the established practices in high-energy physics~\cite{Fruhwith2000}.
Several pattern-recognition algorithms have been applied: for example, the Hough Transform~\cite{Bradt2017,Dalitz2017} and Random Sample Consensus (RANSAC)~\cite{ayyad2018novel,Zamora2021} for straight tracks in three dimensions, as well as approaches based on point triplets~\cite{Dalitz2019} and the Kalman filter~\cite{Ayyad2023} for tracks lacking an analytical description.

In this work, we focus on rectilinear tracks, for which RANSAC is considered the reference method due to its robustness and efficiency.
In RANSAC, a fixed number of points from the recorded set are randomly sampled and checked against a model (in this case a straight line) to determine inliers and outliers according to a chosen threshold.
The sampling is repeated a specified number of times, and the configuration with the largest number of inliers is ultimately selected.

The application of RANSAC to multi-track detection presents several challenges.
First, selecting an appropriate distance threshold is nontrivial, as the actual track width may vary with energy, scattering, or detector resolution, and is often unknown \emph{a priori}.
Second, the close spatial proximity of beam and recoil tracks frequently leads to ambiguous points that could plausibly belong to both structures.
In standard sequential implementations of RANSAC, such points are typically assigned to the first model that detects them and are then removed from the dataset.
This hard assignment prevents those points from informing subsequent model fits, which can lead to suboptimal reconstructions---especially in overlapping regions such as the beam–recoil interface.
Third, low-energy recoils often undergo multiple Coulomb scattering in the detector gas, resulting in curved or kinked trajectories that violate RANSAC’s straight-line model assumptions.

An alternative RANSAC approach involves suppressing the sampling probability of inliers rather than removing them outright.
This allows ambiguous points to reappear in future fits, making the process less rigid than hard removal.
However, this method still lacks a principled weighting scheme and treats shared points as randomly reselectable rather than as simultaneous contributors to multiple models.

In contrast, Gaussian Mixture Models (GMMs)~\cite{dempster1977maximum} assign each point a probabilistic weight for every model component, reflecting the likelihood that it belongs to each track.
This soft assignment enables ambiguous points---such as those near the beam–recoil interface---to influence multiple models simultaneously in a weighted fashion.
Rather than being forced into one model or randomly revisited, such points contribute directly to the estimation of all relevant components.
As a result, GMMs offer a more flexible and accurate reconstruction strategy, particularly in regions with overlapping track geometries.
GMMs are also data-driven: the structure of the tracks is inferred from the point distribution, in contrast to RANSAC, in which a model is imposed on the sampled dataset.
This provides greater flexibility in handling varying track widths and slightly deviated rectilinear shapes.
Despite these advantages, the use of GMMs for track reconstruction in active target detectors has remained largely unexplored.

In this paper, we utilize the \textsc{scikit-learn}~\cite{scikit-learn} implementation of GMMs to distinguish recoil tracks emitted in close proximity to the beam.
To evaluate this approach, we address the following key questions:
(a) For low-energy, short-angled tracks, how effective is the GMM in accurately resolving recoil trajectories?
(b) For high-energy, longer tracks with larger angles relative to the beam, can the GMM be reliably applied?
(c) How well can the method handle holes or missing segments in trajectories, which RANSAC naturally accommodates?
(d) What is the overall efficiency of the GMM-based approach for reconstructing this class of reactions?


\section{Case study: inelastic scattering of \ce{^{58}Ni} on $\alpha$ particles}
\label{case}

To illustrate the performance of the GMM we consider a reaction which was performed at the GANIL facility (Caen, France) in 2019, for the study of isoscalar giant resonances in \ce{^{58}Ni} via inelastic scattering on $\alpha$ particles.
The ACTAR TPC active target~\cite{mauss2019commissioning} was used to acquire the experimental data.
Even though the \ce{^{58}Ni} nucleus is stable, the experiment was conducted in inverse kinematics to characterize the detection setup in view of measurements with unstable nuclei.

\begin{figure}[t]
    \centering
    \includegraphics[width=\textwidth]{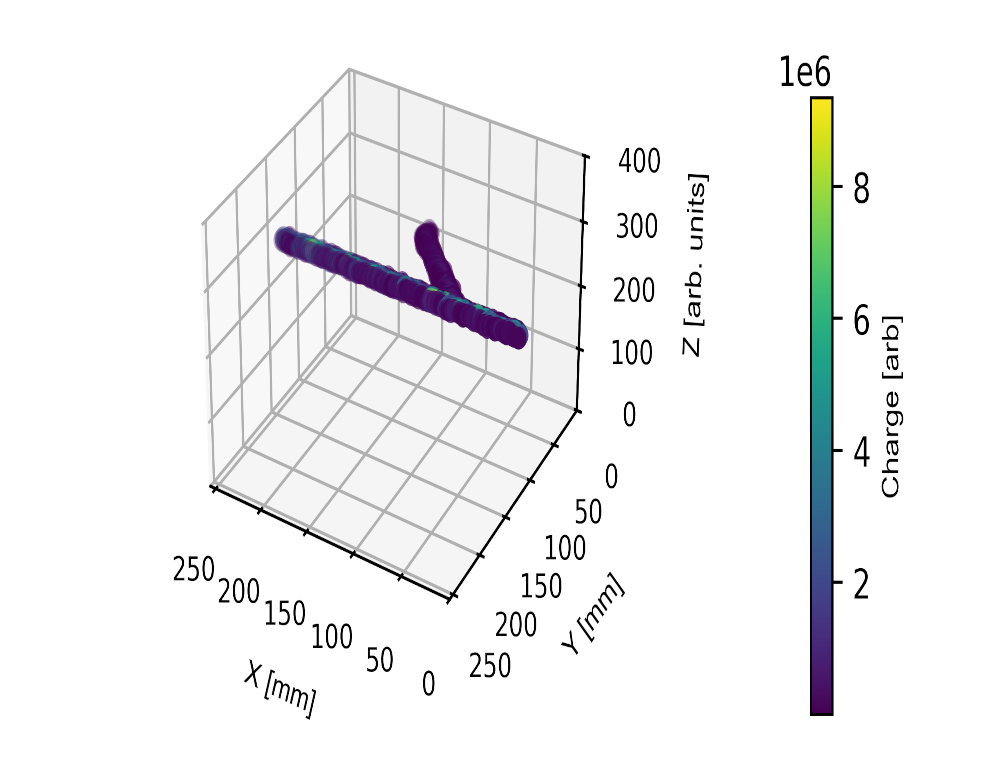}
    \caption{Example event recorded with ACTAR TPC for \Niaa. The beam enters the active volume -- a cubic chamber of dimensions $256 \times 256 \times 256$~mm$^3$ -- at $X = 0$~mm, $Y = 128$~mm, $Z = 128$~mm. The segmented pad plane is located in the $XY$ plane at $Z = 256$~mm. The $Z$ coordinate corresponds to the electron drift time to the segmented plane; values are in arbitrary units.}
    \label{fig:a_fig1}
\end{figure}

The 52-MeV/nucleon beam of \ce{^{58}Ni} nuclei was produced at the LISE facility in GANIL.
The ACTAR TPC active target was filled with a gas mixture of 98\% He and 2\% \ce{CF_4} at 400~mbar.
A uniform electric field was applied in the $256 \times 256 \times 256$~mm$^3$ active volume of the gas target.
Ionisation electrons produced by charged particles traversing the gas were collected on a pad plane, situated above the gas volume, segmented into 16,384 $2\times 2$~mm$^2$ pads.
Amplification of the electron signal was achieved by a Micromegas mesh~\cite{Gio96}.
An example of a recorded event is shown in Fig.~\ref{fig:a_fig1}.

The beam enters the active volume along the $X$-axis, with the segmented plane located above the volume at $Z = 256$~mm.
The $Z$ axis is oriented perpendicular to the segmented plane and represents the drift time axis.
Each recorded event consists of the signal amplitudes and their associated arrival times measured on all pads.
The projection of the signals onto the segmented plane provides two-dimensional information about the track, while the third coordinate is reconstructed from the recorded drift times.
The combined three-dimensional information $\mathbf{x}_i$ defines a \textit{voxel} $i$ within the active volume of the detector.
In order to validate the performance of the proposed model, a simulated data set was generated for the same reaction using NPTool \cite{Matta_2016}, a Geant4-based software.
The segmentation of the pad plane was described using $2\times 2$~mm$^2$ contiguous squares of copper material.
The Micromegas mesh was modeled as a thin aluminum layer placed at 220~$\upmu$m~\cite{mauss2019commissioning} above the pad plane (amplification gap).
The Geant4 field manager was used to create the constant electric field ($70$~V/cm).
The electromagnetic physics constructor \textsc{emstandard} in Geant4 was used to handle ion tracking.
The number of ionization electrons created in each simulation step was determined by the energy deposited by the charged particle traversing the gas, divided by the ionization potential of helium.
The transport parameters for the electrons in the gas mixture were taken from experimental data.
The drift velocity was determined to be $v_\mathrm{d} = 1.16~$cm/$\upmu$s, by measuring the time difference between the arrival of electron signals generated by the same track at known points of the volume (typically the beam position and the edge of the detection volume).
The transverse ($\sigma_\mathrm{T}$) and longitudinal ($\sigma_\mathrm{L}$) spreads in the velocity vector were calculated as $\sigma_x = \sqrt{2\,S_x/\Delta t}$ ($x=\mathrm{L,T}$), where $\Delta t$ is the time step of the simulation of the drifting electron and $S_\mathrm{T}$ and $S_\mathrm{L}$ are the transverse and longitudinal diffusion coefficients, respectively.
The parameters used in the simulation are $S_\mathrm{T} = 2\times 10^{-5}$~mm$^2$/ns and $S_\mathrm{L} = 4\times 10^{-6}$~mm$^2$/ns, which reproduce the spread in the experimental data.
Geant4 secondary electron production cuts were also optimized to generate tracks that have a similar width profile as the tracks in the experimental data.

The simulated data set used in this study corresponds to a fixed excitation energy of 10\,MeV for the $^{58}$Ni nucleus and spans center-of-mass scattering angles $\theta_\mathrm{c.m.}$ from $1^\circ$ to $5^\circ$ in 1-degree increments.
These angles correspond to increasing recoil energies and laboratory-frame angles: scattering at $1^\circ$ in the center-of-mass reference frame translates to a laboratory angle of $32^\circ$ and a recoil energy of approximately 0.18~MeV (corresponding to a track length of $\sim$30\,mm in the gas), while $\theta_\mathrm{c.m.}=5^\circ$ maps to an angle in the laboratory frame of $70^\circ$ and a recoil energy of approximately 1.43~MeV ($\sim$116\,mm in the gas).
This selection provides a representative spectrum from short, low-energy tracks at shallow angles with respect to the beam direction to longer, higher-energy tracks at steeper angles.
This enables us to assess our implementation of the GMMs across a range of physically relevant conditions.


\section{Track reconstruction using the Gaussian Mixture Model}
\label{track_reconstruction}

Track reconstruction proceeds through three stages: 1) cluster identification, 2) cluster regularization, 3) determination of reaction vertex and track length.

\subsection{Identification of Clusters}
\label{track_identification}

\begin{figure}[t]
	\centering
	\medskip
	\includegraphics[width=1.0\textwidth]{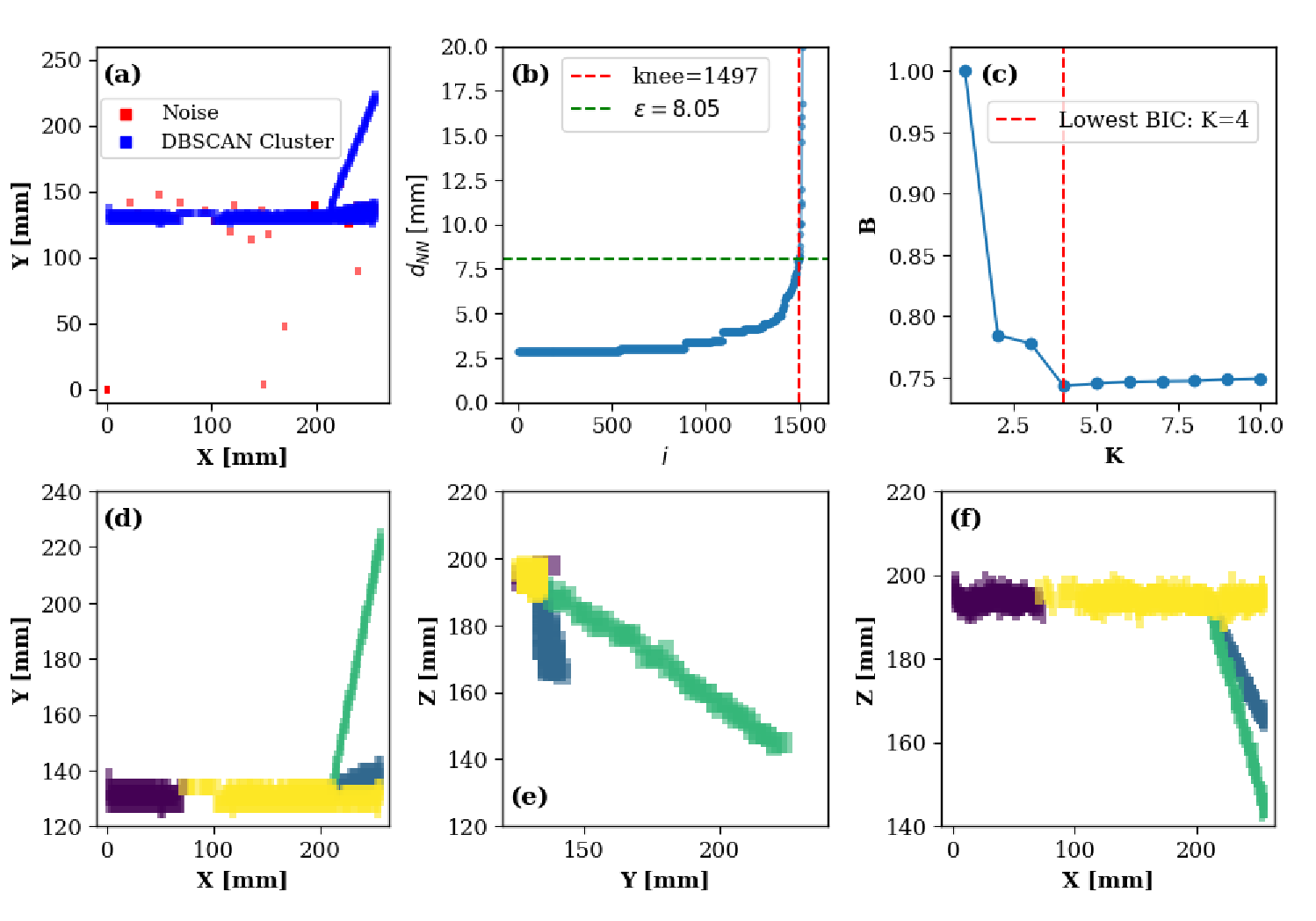}
	\caption[Short caption for Table of Figures]{An event from the experimental data of the \Niaa\ reaction, as recorded in ACTAR TPC. (a) Cluster identified using the DBSCAN algorithm is shown in blue, excluded voxels are shown in red (only one cluster is identified in this event); (b) Fifth Nearest neighbour distance $d_\mathrm{NN}$ in mm (sorted), measured for each voxel $i$ recorded in the event; (c) Evaluation of the normalized Bayesian Information Criterion (BIC) score $B(K)$ 
	as a function of the number of components $K$; (d), (e), and (f): respectively $XY$, $YZ$, and $XZ$ projections of the $K=4$ clusters identified using the GMM; each color correspond to a single identified component $k$. We recognize the beam track, identified as two separate tracks (purple and yellow) by the GMM, and two scattered tracks.} 
	\label{fig:b_fig2}
\end{figure}

As a first step in the identification of the tracks, regions in the active volume with a high density of recorded signals are identified using the DBSCAN algorithm \cite{Ester1996density}.
This algorithm takes two parameters as input: a distance $\epsilon$ and minimum number of voxels $n_\mathrm{min}$.
The algorithm selects \emph{core voxels} if they have at least $n_\mathrm{min}$ voxels within the distance $\epsilon$.
Voxels that do not meet this criterion, but lying within $\epsilon$ from a core voxel, are classified as \emph{border} voxels and form a part of the identified cluster.
Fig.~\ref{fig:b_fig2}a shows the result for an event recorded in ACTAR TPC (note that, in some cases, more than one DBSCAN cluster may be identified in an event).

The $n_\mathrm{min}$ parameter is set according to the standard recommendation \cite{schubert2017dbscan} ($n_\mathrm{min}=2\times dim=6$, where $dim$ is the number of dimensions). The distance $\epsilon$ is calculated from the knee point of the graph showing the fifth nearest-neighbour distance ($d_{NN}$, chosen according to \cite{schubert2017dbscan} = (2 $\times$ $dim-1$) for each voxel recorded in the event\footnote{The knee point is calculated for each event using functions available in \textsc{scikit-learn} \cite{scikit-learn}.}, as shown in Fig.~\ref{fig:b_fig2}b.

Since the DBSCAN algorithm groups all densely connected voxels into one cluster, the ejectile tracks cannot be distinguished from the beam tracks, see Fig.~\ref{fig:b_fig2}a.
The GMM is then used to re-cluster the voxels in each DBSCAN cluster according to their conformity to 3-dimensional cylindrical shapes.
More precisely, every voxel $\mathbf{x}_i$ is assumed to be drawn from a mixture of $K$ 3-dimensional gaussian components (eventually identified as particle tracks) represented by 
\begin{equation}
p(\mathbf{x}_i) = \displaystyle\sum_{k=1} ^{K} \pi_{k} \mathcal{N}(\mathbf{x}_i|\bm{\upmu}_k,\bm{\Upsigma}_k)
\end{equation}
where the mixing coefficient $\pi_k$ is the prior ($\sum_{k=1}^{K} \pi_k = 1$, $\pi_k \geq 0$) and the conditional probability for drawing a voxel $\mathbf{x}_i$ from a component $k$ is $\mathcal{N}(\mathbf{x}_i|\bm{\upmu}_k,\bm{\Upsigma}_k)$, with $\bm{\upmu}_k$ and $\bm{\Upsigma}_k$ representing the (three-dimensional) mean and the covariance of the $k^\mathrm{th}$ component.
The posterior probability ($\gamma_{ki}$) for a component $k$ to have generated the voxel $\mathbf{x}_i$ is then given by 
\begin{equation}
\gamma_{ki} = \frac{\pi_{k} \mathcal{N}(\mathbf{x}_i|\bm{\upmu}_k,\bm{\Upsigma}_k)}{\displaystyle\sum_{j=1} ^{K} \pi_{j} \mathcal{N}(\mathbf{x}_i|\bm{\upmu}_j,\bm{\Upsigma}_j)}
\end{equation}
The maximum likelihood solutions are found by the Expectation-Maximization approach in three steps~\cite{Bishop2006_Ch9}: 
\begin{enumerate}
	\item the means $\bm{\upmu}_k$ and covariances $\bm{\Upsigma}_{k}$ of the 
	components are initialized using the $K$-means algorithm \cite{hartigan1975clustering}.
    $\pi_k$ is initialized with values proportional to the number of voxels in each component $k$.
    The covariance type is chosen to be fully allowing non-zero off-diagonal elements and non-equal diagonal elements.
    This allows the model to find line-like 3-dimensional structures; 
	\item the posterior probabilities $\gamma_{ki}$ of each component for all the voxels 
	are evaluated for the assumed parameters $\pi_k$, $\bm{\upmu}_k$ and $\bm{\Upsigma}_k$;
    each voxel $\mathbf{x}_i$ is (re-)assigned to the cluster $k$ for which $\gamma_{ki}$ is maximum; 
	\item the parameters $\pi_k$, $\bm{\upmu}_k$ and $\bm{\Upsigma}_k$ are re-evaluated with the new $k$ components.
\end{enumerate}
The procedure is repeated until convergence (no large changes in $\pi_k$, $\bm{\upmu}_k$ and $\bm{\Upsigma}_k$ based on a specific tolerance).

In the kinematic context, each identified component $k$ is synonymous with a track.
The value of $K$ (total number of clusters) cannot be chosen too small because, in some experimental events, multiple tracks are present in a single DBSCAN cluster.
An optimum value of $K$ can be evaluated for each event, by carrying out the procedure described above for values of $K$ from 1 to a maximum value $K_\mathrm{max}$ (we used $K_\mathrm{max}=10$) and by using the Bayesian Information Criterion (BIC) \cite{Schwarz1978}.
For each value of $K$ a normalized BIC score $B(K)$ is calculated:
\begin{align}
B(K) &= 
 -2 \sum_{i=1}^{N} \ln \left( \sum_{k=1}^{K} \pi_k \mathcal{N}(\mathbf{x}_i | \bm{\upmu}_k, \bm{\Sigma}_k) \right) + (10K-1) \ln(N)
\label{eq:bic_score}
\end{align}
where $N$ is the number of voxels in the original DBSCAN cluster.
The normalized BIC score $B(K)$ as a function of the number of components $K$ is shown in Fig.~\ref{fig:b_fig2}c.
According to the criterion, the number of components ($K$) with the lowest $B(K)$ is chosen ($K=4$ in the figure).
In Fig.~\ref{fig:b_fig2}d-f, the components identified with the GMM are shown with different colors.

\begin{figure}[t]
	\centering
	\medskip
	\includegraphics[width=1.0\textwidth]{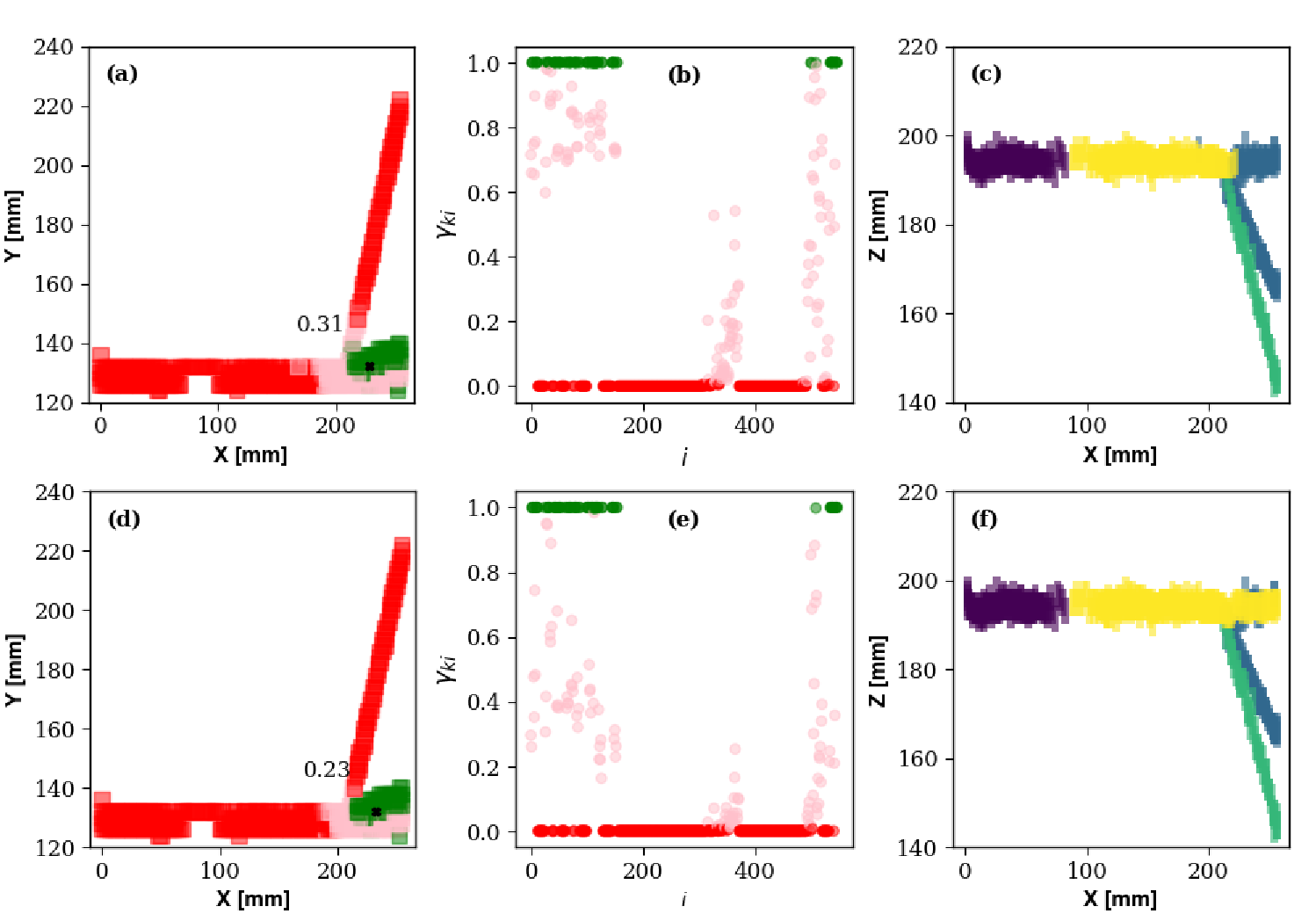}
	\caption[Short caption for Table of Figures]{Assignment of posterior probabilities $\gamma_{ki}$ of voxels $i$ 
	to one component $k$ --- the blue component in (c) and (f) --- in successive iterations.
    (a) projection of voxels $i$ 
    on the $XY$ plane, color-coded by their probabilistic assignment $\gamma_{ki}$: red are excluded voxels ($\gamma_{ki}$ = 0), green are assigned voxels ($\gamma_{ki}$ = 1), pink are ``soft-assigned'' voxels ($0 < \gamma_{ki} < 1$). The black cross marks the assigned mean vector $\bm{\upmu_k}$ and the number is the prior $\pi_k$; (b) posterior probabilities ($\gamma_{ki}$) for each voxel $i$; 
    (c) identified color-coded components;
    (d), (e), (f): same as (a), (b), (c) but for the successive iteration.} 
	\label{fig:c_fig3}
\end{figure}

Fig.~\ref{fig:c_fig3} shows, for this same event and for the chosen value $K=4$, some steps of the procedure for the identification of the GMM clusters.
Fig.~\ref{fig:c_fig3}a,d show the assignment of posterior probabilities $\gamma_{ki}$ of each voxel $i$ to a specific component $k$.
The assignments are categorized into ``excluded'' ($\gamma_{ki} = 0$, red color), ``soft'' ($0 < \gamma_{ki} < 1$, pink) and ``full'' ($\gamma_{ki} = 1$, green).
The evolution of these assignments in successive iterations is shown in Fig.~\ref{fig:c_fig3}b,e.
The newly generated components $k$ based on the probabilistic assignment of the voxels is shown in Fig.~\ref{fig:c_fig3}c,f.
One can follow the blue-colored component in Fig.~\ref{fig:c_fig3}c,f to trace the evolution of the voxel assignments near the beam-ejectile interface.

An important feature of this method is the possibility of applying different criteria to assign a voxel $\mathbf{x}_i$ to a cluster $k$, based on the value of the posterior probability $\gamma_{ki}$.
For example, initially a voxel can be assigned to the cluster for which $\gamma_{ki}$ is largest.
In successive steps, a threshold on $\gamma_{ki}$ can be imposed, allowing a voxel to be assigned to more than one cluster.
This accounts for the fact that the charge in a voxel may originate from multiple tracks, which is especially relevant in regions where a scattered track overlaps with the beam.

\subsection{Cluster regularization}
\label{cluster regularization}

In some cases the GMM leads to overfitting, i.e., a large number $K$ of identified tracks, when the increase in the log-likelihood outweighs the penalty for high $K$ in equation~\ref{eq:bic_score}.
Some of the reasons may be: i) the recorded voxels do not conform to the assumed cylindrical shapes, with high variance in one direction and minimal variance in the perpendicular directions; this may occur when low-energy tracks are scattered in the medium, resulting in non-rectilinear trajectories; ii) non-working pads may lead to the initial identification of multiple DBSCAN clusters in an event; and iii) in a small DBSCAN cluster, the low value of the number of voxels $N$ would weaken the penalty $\ln(N)$ for higher $K$.

\begin{figure}[t]
	\centering
	\medskip
	\includegraphics[width=1.0\textwidth]{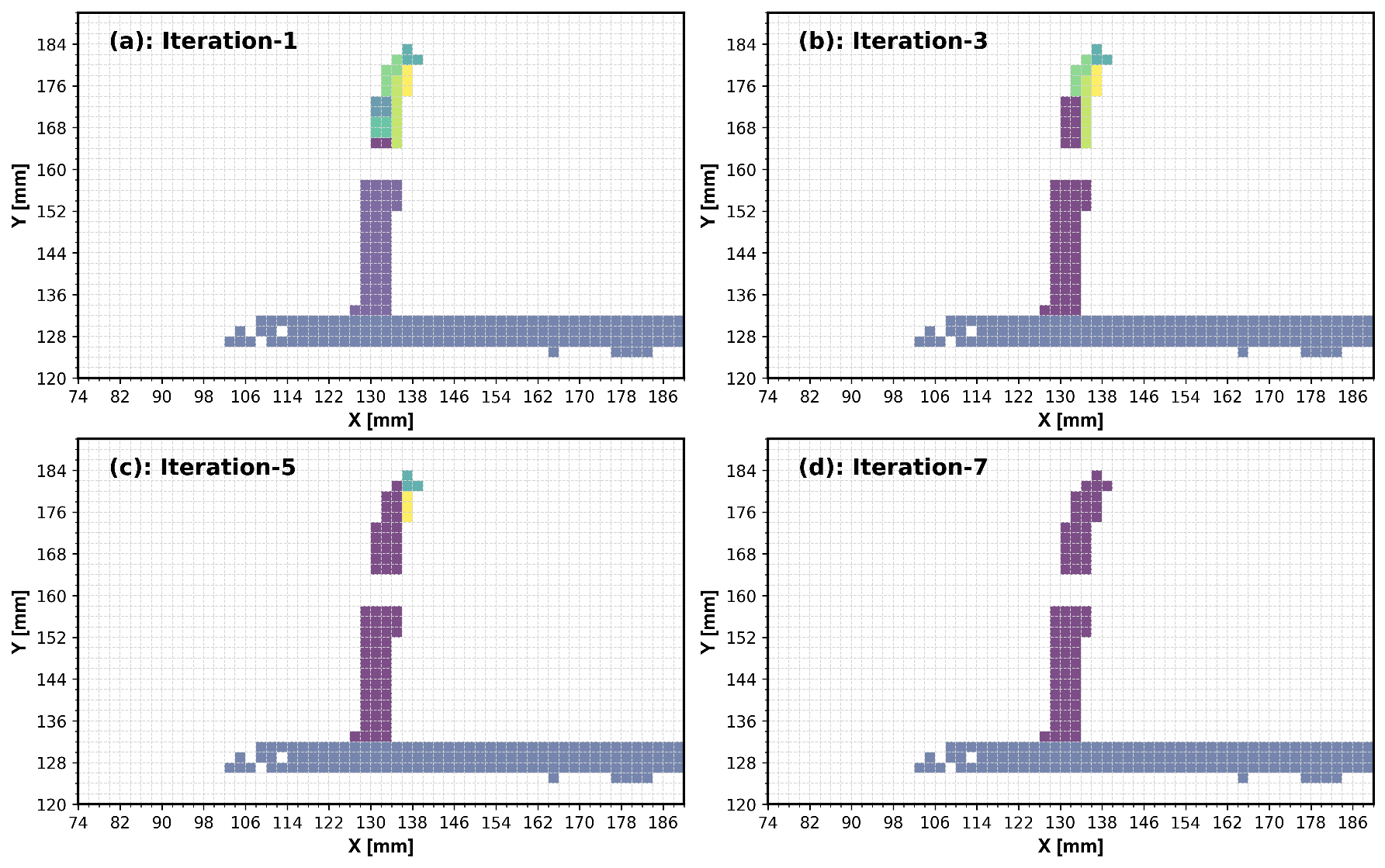}
	\caption[Short caption for Table of Figures]{Clusters from a \Niaa\ event recorded in ACTAR TPC, color-coded by their component $k$. (a) Assigned components by GMM; (b),(c) Re-assignment after a certain number of iterations; (d) Assigned components after regularization.}
	\label{fig:d_fig4}
\end{figure}

Fig.~\ref{fig:d_fig4}a shows the clusters generated by the GMM in an event in ACTAR TPC, in which the overfitting problems are visible.
The gap around $Y\approx 160$~mm is the result of non-working pads.
In the region from $Y\approx 160$~mm to 180~mm, the GMM identifies several clusters.
These issues can be mitigated through a regularization procedure, aiming at reducing the number of clusters $K$ identified in an event by the GMM, according to specific merging criteria.
In the procedure, an appropriate metric $m_{kl}$ is constructed between each pair of clusters; if the metric exceeds a threshold $m_\mathrm{th}$, which is determined phenomenologically, the clusters are merged.

This is implemented step-wise: at each step, a cluster is merged with (at most) one other cluster, for the pair which has the largest $m_{kl}$ (and it must be $m_{kl}>m_\mathrm{th}$).
The comparison is iterated until the remaining clusters do not satisfy the metric criteria.

The choice of metric depends upon the issue to be regularized.
To merge GMM clusters which are aligned in three dimensions, an effective criterion is the average 3-dimensional Mahalanobis distance \cite{Mahalanobis18} between a cluster $k$ and the $N_l$ points of a cluster $l$.
More specifically, one can use the $p$-value of the distance\footnote{A large $p$-value means the observed separation (Mahalanobis distance) is not statistically significant,  i.e. the two clusters are consistent with coming from the same underlying track and therefore can be merged.}:
\begin{equation}
\label{pij_Mahalanobis_distance}
p_{kl} = 1-\chi^2\left[\frac{\sum_{i=1}^{N_l}\sqrt{{(\mathbf{x}_i-\bm{\upmu}_k)^\intercal\, \bm{\Upsigma_k}^{-1}\,(\mathbf{x}_i-\bm{\upmu}_k})}}{N_l}, 2\right]
\end{equation}
Fig.~\ref{fig:d_fig4}b-d show the progression at different steps of the regularization procedure when using this metric.
The threshold should be chosen such that the tracks of ejectiles are reconstructed correctly, while still remaining separate from the beam tracks.
To estimate the value of the threshold, the $p_{kl}$ are calculated for the simulated data, where one has prior knowledge of the nature of the tracks.

\begin{figure}
	\centering
	\medskip	
	\includegraphics[width=0.9\textwidth]{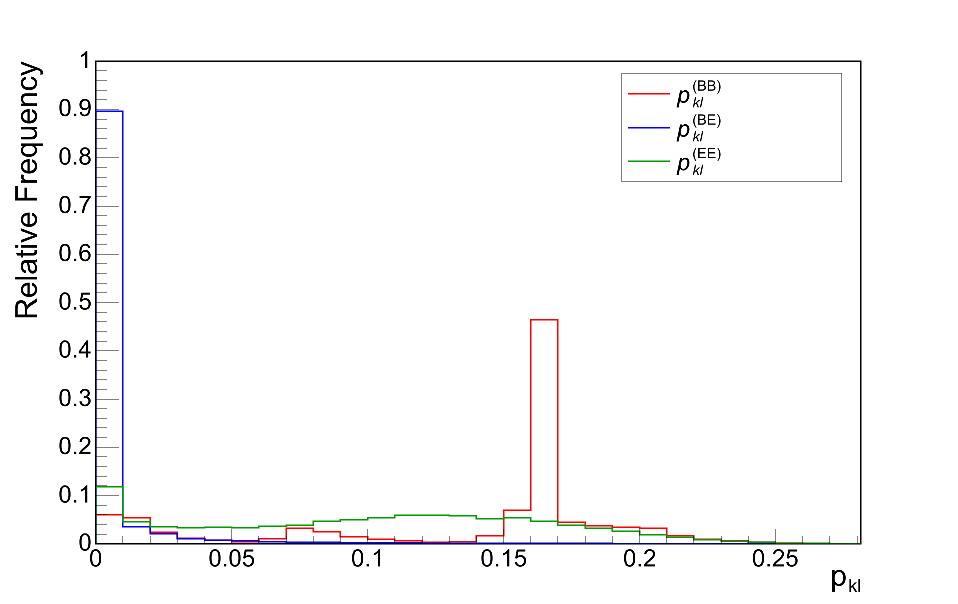}
	\caption[]{Relative frequency of the pairwise metric $p_{kl}$ computed between all distinct pairs of predicted GMM tracks $K$ in an event, for the entire simulated data set. Three categories of the metric are shown: Beam-Beam Metric \( p_{kl}^{(BB)} \), Ejectile-Ejectile Metric \( p_{kl}^{(EE)} \), Beam-Ejectile Metric \( p_{kl}^{(BE)} \).}
	\label{fig:e_fig5}
\end{figure}

Fig.~\ref{fig:e_fig5} shows the $p_{kl}$ values calculated on the simulation, separately for pairs of beam tracks ($p_{kl}^\mathrm{BB}$), pairs of ejectile tracks ($p_{kl}^\mathrm{EE}$), and pairs formed by one beam track and one ejectile track ($p_{kl}^\mathrm{BE}$).
We clearly see that the $p$-values are very small in the case of beam-ejectile pairs, while they are grouped at higher values for beam-beam pairs.
By choosing, for example, a threshold equal to 0.1, we ensure that the procedure does not merge the beam and ejectile tracks together, while it correctly merges the beam-like tracks.

\begin{figure}
	\centering
	\medskip	
	\includegraphics[width=0.9\textwidth]{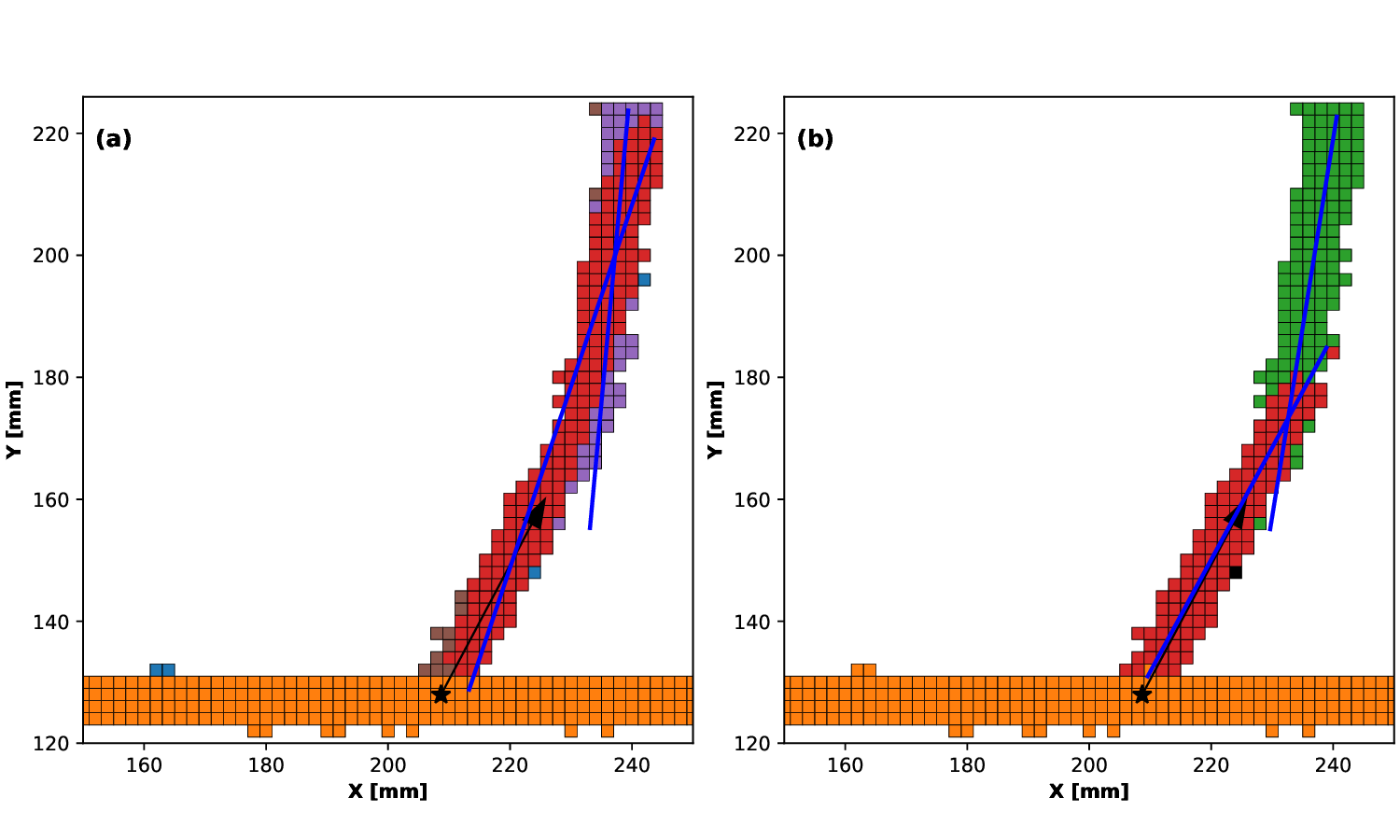}
	\caption[]{(a) Clusters colored according to the RANSAC classification with a distance threshold of 5 mm. (b) Clusters colored according to Gaussian Mixture Model (GMM) components regularized by $p_{kl}$. The black star marks the simulated reaction vertex, while the black arrow indicates the simulated direction vector of the ejectile. The blue lines represent the fitted direction vectors (see section \ref{krecon}) of the reconstructed track segments}
	\label{fig:f_fig6}
\end{figure}

The $p$-value for pairs of ejectile tracks is more widely distributed, due to possible scattering at low energy that changes the direction of the ejected particle, resulting in a ``broken'' or ``segmented'' track.
Such effects can be especially pronounced when the detection gas contains atoms which are heavier than the ejectile particles.
In RANSAC, the correct identification of such events depends crucially upon the choice of the threshold distance that selects inliers and outliers.
A large threshold may lead to the inclusion of all points into one rectilinear track, but at the expense of the correct derivation of the scattering angle and the range of the particle.
A smaller threshold, on the other hand, may lead to the identification of multiple tracks and the subsequent rejection of the event when the analysis requires only one ejectile track.
Figure~\ref{fig:f_fig6} illustrates such a representative case.
This figure demonstrates how the GMM approach preserves the segmentation information of a deviated track, in contrast to the RANSAC method, where such information is lost.

\begin{figure}
	\centering
	\medskip	
	\includegraphics[width=1.0\textwidth]{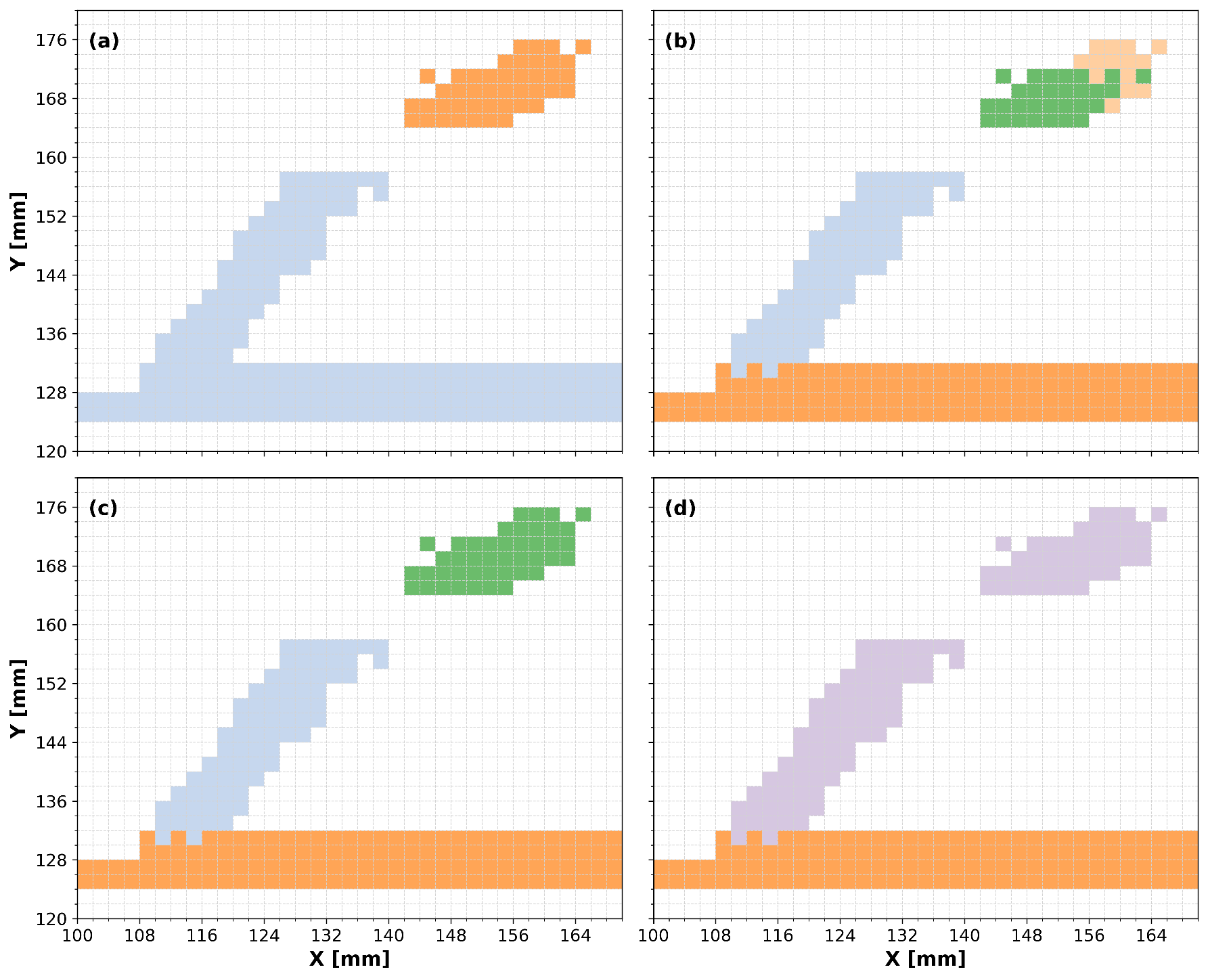}
	\caption[]{Cluster assignments for an event from the simulated dataset. (a) Clusters labeled according to DBSCAN. (b) Clusters labeled according to GMM. (c) Clusters labeled after the GMM clusters are regularized using the $p_{kl}$ metric. (d) Clusters labeled after additional regularization using the $c_{kl}$ metric. Each panel shows the XY projection of the voxels $i$, with colors representing the cluster labels.}
	\label{fig:g_fig7}
\end{figure}

To properly take these effects into account, another regularization procedure with an appropriate metric can be implemented.
For every segment in the GMM clusters regularized by $p_{kl}$, the two extremities $\mathbf{s}_k$ (start, closest to the middle of the chamber) and $\mathbf{e}_k$ (end) are identified.
The metric is defined as
\begin{equation}
    c_{kl} = \min(||\mathbf{e}_k-\mathbf{s}_l||,||\mathbf{e}_l-\mathbf{s}_k||)
\end{equation}
With this choice\footnote{The term $min(||\mathbf{e}_k-\mathbf{e}_l||)$ is excluded, since it would approximate the behavior of a larger RANSAC threshold.}, two tracks are merged when they form a continuous pattern, even if the direction changes.
Figure~\ref{fig:g_fig7} summarizes the clustering workflow: (a) initial DBSCAN cluster, (b) corresponding GMM clusters, (c) regularization using $p_{kl}$, and (d) further refinement using $c_{kl}$.

By setting a threshold, tracks as in Fig.~\ref{fig:g_fig7}, for which the deviation coincides with an area of non-working pads, can also be regularized.
The choice of the threshold in this case is guided by the presence of such non-working areas and their size.

\subsection{Reaction vertex, track length, and scattering angle}
\label{krecon}

In the analysis of inelastic scattering data, kinematic reconstruction is employed to determine the excitation energy \(E^*\) of the heavy projectile and the center-of-mass scattering angle \(\theta_\mathrm{c.m.}\).
These quantities are extracted from the experimental observables: the laboratory scattering angle \(\theta_\mathrm{lab}\) (defined as the angle between the beam track and the ejectile track), and the ejectile energy \(E_{\alpha}\), which is inferred from the measured track length.
Further in this paper, we will use these quantities to compare the performance of the GMM to the RANSAC method.
Here, we briefly illustrate the procedures that we adopted to extract them.

A first step consists of the identification of the ejectile track and the beam track.
In ACTAR TPC, the beam zone is defined as the region $Y= 122$ to 132~mm; this is also referred to as the \emph{attenuation} zone because, experimentally, a reduced amplification is applied in this area of the pad plane to avoid saturation.
Tracks whose centroid $\bm{\upmu}_k$ falls within the attenuation zone are classified as beam tracks, while others are classified as ejectile tracks.

\begin{figure}[t]
	\centering
	\medskip	
	\includegraphics[width=1.0\textwidth]{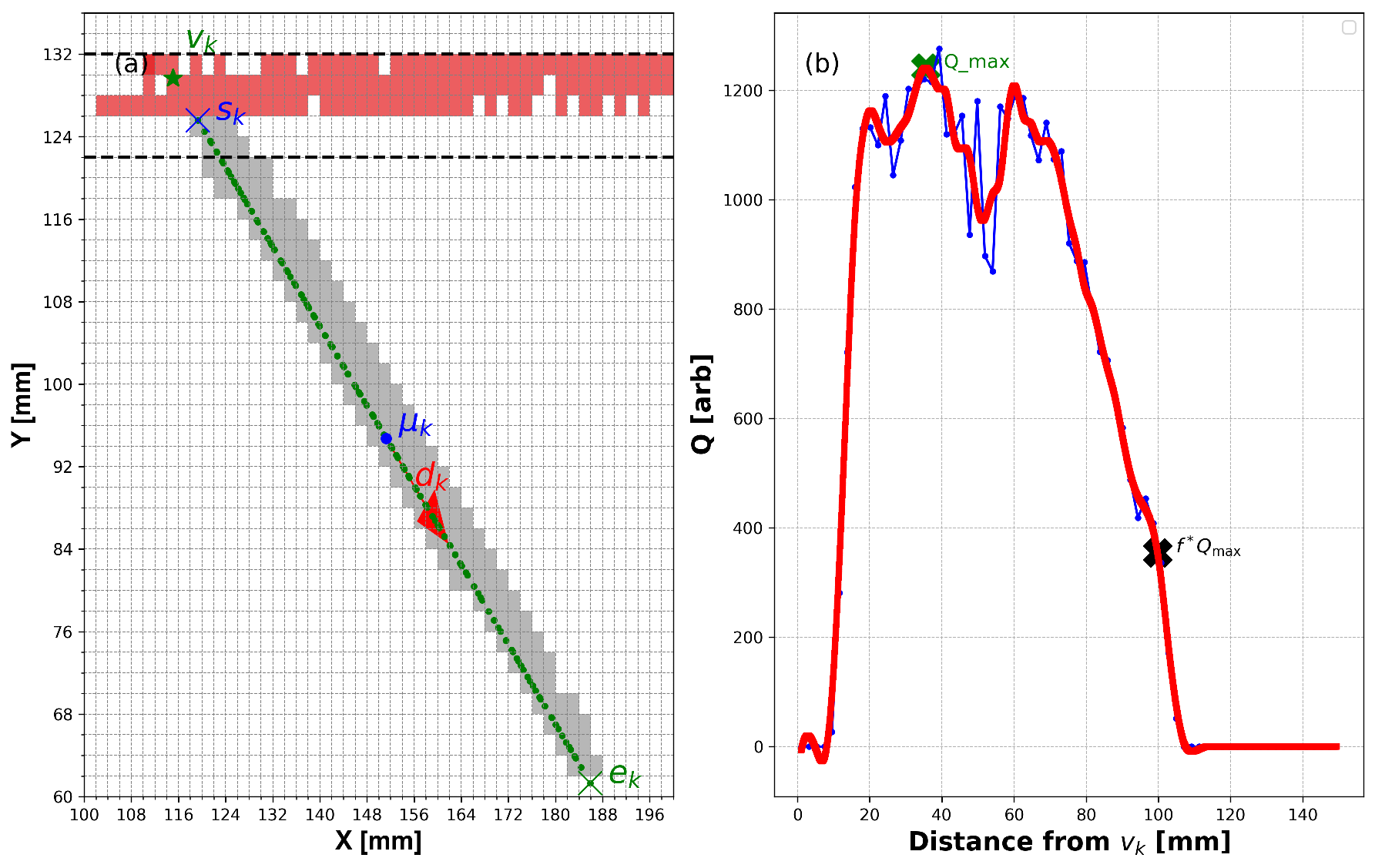}
	\caption[]{Determination of the reaction vertex ($\mathbf{v}_k$) and the track length ($R$) for an event in ACTAR TPC. (a) The GMM identifies the beam track (red voxels in the attenuation zone, indicated by the dashed black lines) and the ejectile track (grey voxels); on the latter, the points closest to each voxel (green dots) along the direction $\hat{\mathbf{d}}_k$ (first principal component, red arrow) are shown; also along $\hat{\mathbf{d}}_k$, the start $\mathbf{s}_k$, middle $\bm{\upmu}_k$, and end $\mathbf{e}_k$, as well as the reaction vertex $\mathbf{v}_k$, are identified;
    (b) The energy loss profile of the scattered track along the direction vector $\mathbf{d}_k$. The blue line represents the weighted sum of the charges ($Q$) deposited in each bin, projected along $\mathbf{d}_k$. The red line represents the smoothed energy loss profile obtained using a Savitzky-Golay filter \cite{press1990savitzky} with a window size of 7 mm. The green and black crosses indicate the location of the maximum energy loss ($Q_{max}$) and the point where the energy loss is a fraction $f$ of the maximum, respectively.}
	\label{fig:h_fig8}
\end{figure}

For each ejectile track, the direction $\hat{\mathbf{d}}_k$ is defined as the first principal component of the corresponding cluster, calculated using weighted principal component analysis \cite{delchambre2015weighted}.
The reaction vertex $\mathbf{v}_k$ is the point, on the direction vector $\hat{\mathbf{d}}_k$, closest to the first principal component of the corresponding\footnote{Experimentally, one event may show multiple beam tracks that appear at different $Z$ heights depending on the difference in timing from the trigger of that event. The ``corresponding'' beam track is the one for which the distance to $\mathbf{v}_k$ is the shortest.} beam track.
Fig.~\ref{fig:h_fig8}a shows the parameters $\bm{\upmu}_k$, $\hat{\mathbf{d}}_k$, and $\mathbf{v}_k$ calculated for an event recorded in ACTAR TPC.

The determination of the length of the track is based on the profile of the charge collected along the track.
To calculate the charge profile, a cylinder is defined, along the track direction $\hat{\mathbf{d}}_k$, that contains all the voxels of the cluster.
The cylinder is then sliced into short bins (in our case 2~mm), and the charge $Q$ deposited in each bin is calculated from the sum of the charge deposited in the pads covered by the projection of the bin, weighted by the fraction for which each pad is covered.
The obtained profile is then smoothed by applying a Savitzky-Golay filter \cite{press1990savitzky} with a window size larger than the bin size (in our case, 7~mm)\footnote{This value could be adapted when experimental tracks are very short.}. 
The calculated energy loss profile and the smoothed energy loss profile are shown in Fig.~\ref{fig:h_fig8}b.
The next step is the determination of the point at which the energy loss is maximum ($Q_\mathrm{max}$), indicated by the green cross in Fig.~\ref{fig:h_fig8}b.
Finally, the length of the track $R$ is defined as the distance between $\mathbf{v}_k$ and the point where the collected charge is a chosen fraction $f$ of $Q_\mathrm{max}$.
The value of $f$ is optimized by applying the procedure to simulated data, and by minimizing the differences in the reconstructed kinematic parameters with respect to the input of the simulation.
The ejectile energy $E_{\alpha}$ is derived from the track length by using energy loss tables such as those provided by SRIM \cite{ziegler2010srim}.

\begin{figure}
	\centering
	\medskip	
	\includegraphics[width=0.8\textwidth]{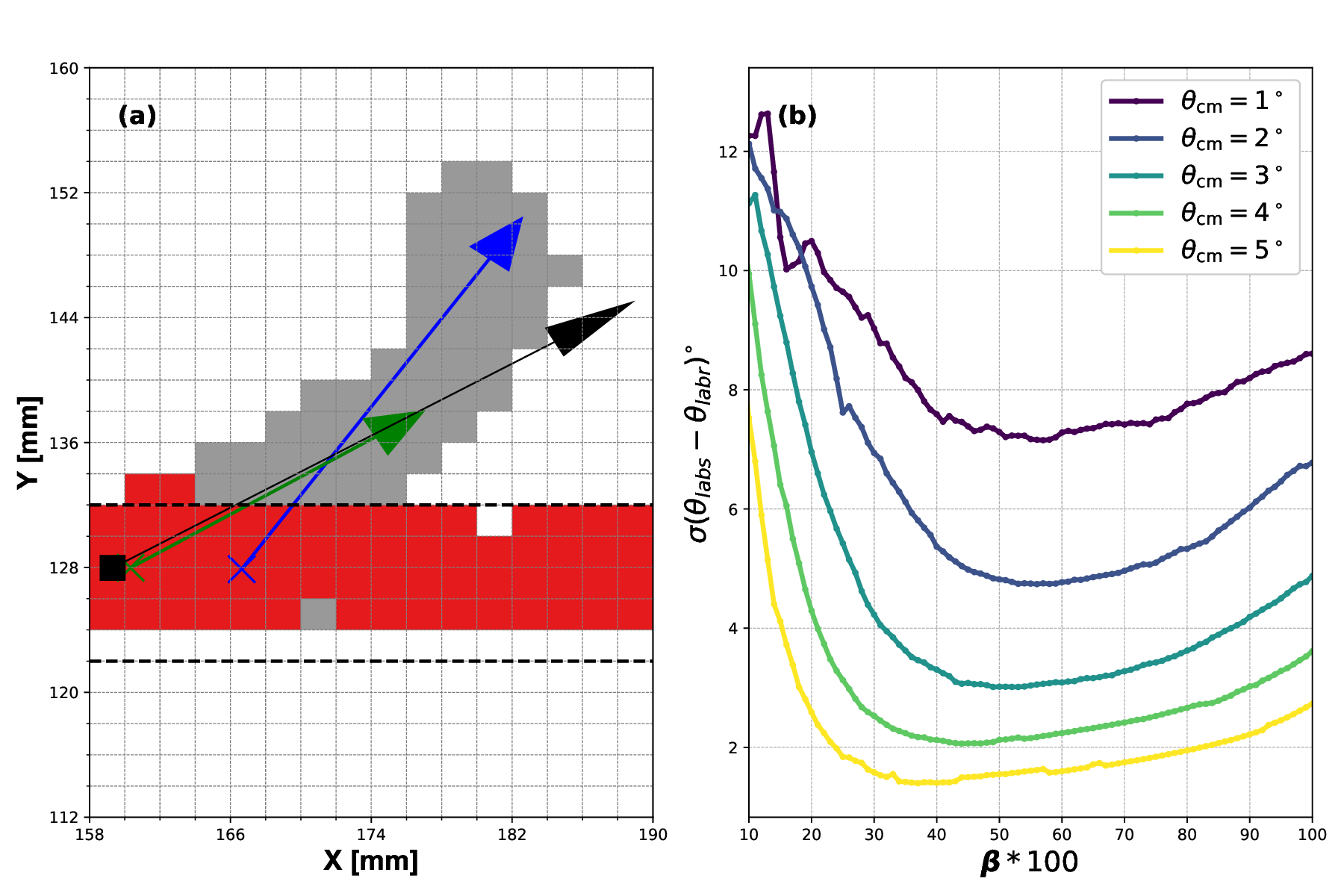}
	\caption[]{Optimization of $\theta_{lab}$ via $\beta$-tuning. (a) The blue cross and the blue arrow correspond to the reconstructed reaction vertex and the direction vector obtained by using the full length of the track. The green cross and the green arrow represent those same quantities, but obtained using the fraction $\beta$ of the total length of the track. The filled black square and the black arrow represent the simulated reaction vertex and the simulated direction vector; (b) standard deviation of the difference between the simulated ($\theta_\mathrm{labs}$) and reconstructed ($\theta_\mathrm{labr}$) lab angle.}
	\label{fig:i_fig9}
\end{figure}

The laboratory angle $\theta_\mathrm{lab}$ is the angle between the direction of the ejectile $\hat{\mathbf{d}}_k$ and the first principal component of the corresponding beam track.
When the first principal component is computed using all points associated with the ejectile track, the true direction of the track may be misrepresented.
This happens, as we have seen above, in cases where the track is distorted or fragmented due to secondary interactions or reconstruction artifacts.
Another example is shown in Fig.~\ref{fig:i_fig9}a.
To improve the determination of the scattering angle, two correction methods are employed.
Firstly, only part of the ejectile track is considered, by keeping just the voxels closest to the beam track, corresponding to a fraction $\beta$\footnote{An alternative approach would be to use a fixed length instead of the fraction. Such an approach could better handle very short tracks.} 
of the length of the track.
This step can be also employed when the tracks are identified by other pattern-recognition methods, such as RANSAC.
Only the GMM method, however, allows using a second technique, by exploiting the knowledge given by the posterior probability \( \gamma_{ki} \) for a voxel to belong to one or the other cluster. 
This way, we can add to the ejectile track the voxels in the attenuation zone for which the posterior probability of belonging to an ejectile track $k$ exceeds a certain threshold \( \gamma \).

This unique feature of the GMM reflects an actual physical situation: charges within the same voxel may indeed be generated (directly or indirectly) by both the beam and ejectile particles.
Inclusion of those voxels in the track of the ejectile particle should lead to an improvement in the extracted scattering angle.

\begin{figure}
	\centering
	\medskip	
	\includegraphics[width=0.8\textwidth]{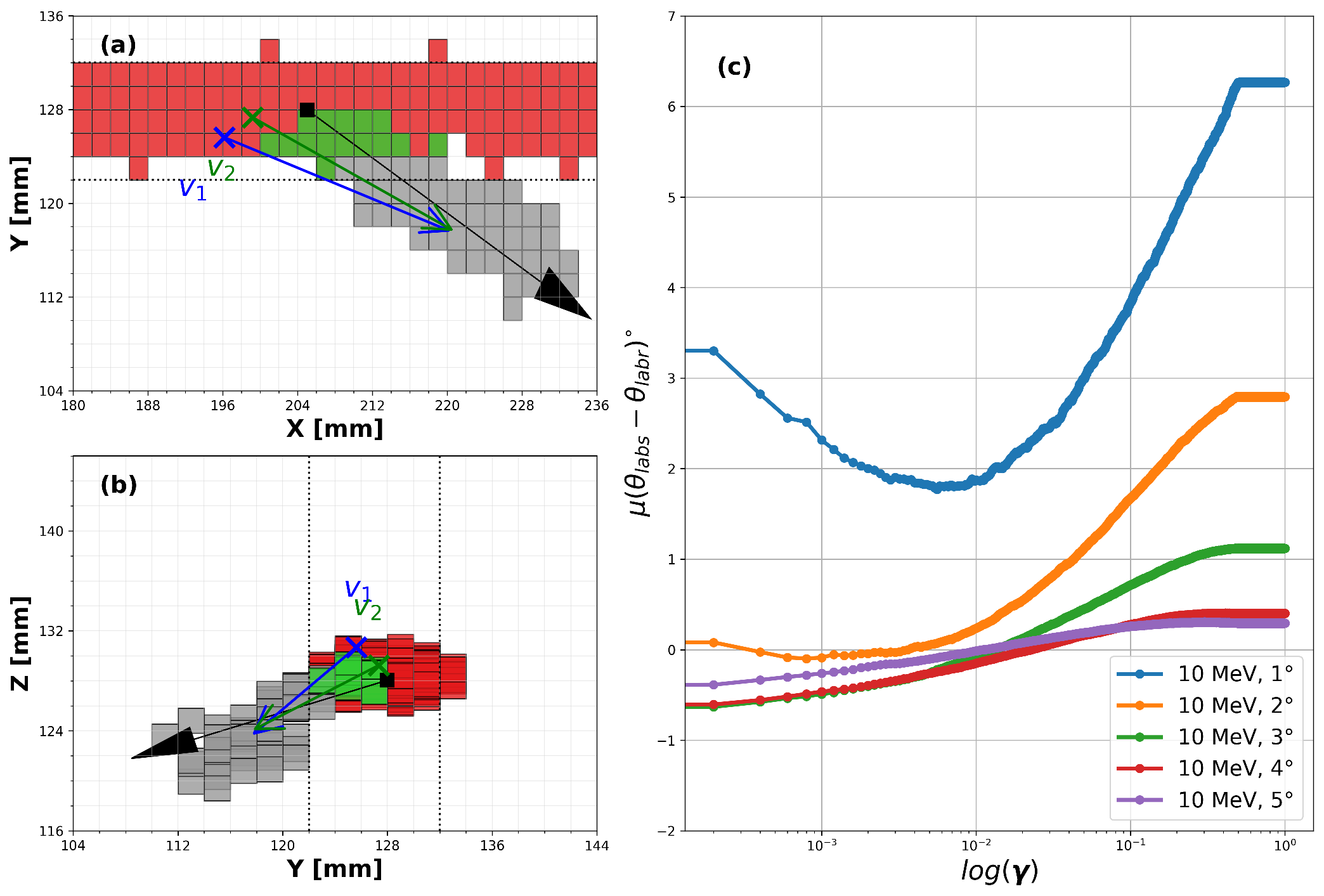}
	\caption[]{(a) Illustration of a representative event (top view, in the $XY$ plane). The voxels assigned to the beam track are shown in red, those of the ejectile track are shown in grey. The blue cross and blue arrow correspond to the initial reconstruction using a fixed fraction \( \beta \) of the total length of the ejectile track. Green pixels indicate voxels initially assigned to the beam track, with posterior probability \( \gamma_{ki} > 1\% \) to belong to the ejectile track. The green cross and green arrow show the second refinement after applying the threshold. The black square and black arrow represent the simulated reaction vertex and simulated direction vector, respectively. (b) Same as (a), but in the (front view) $YZ$ plane. (c) Mean of the difference between the simulated laboratory angle \( \theta_{\text{labs}} \) and the reconstructed laboratory angle \( \theta_{\text{labr}} \) as a function of the threshold \( \gamma \).}
	\label{fig:j_fig10}
\end{figure}

Once more, both the \( \beta \) and \( \gamma \) parameters are determined by applying the method to simulated data and comparing the reconstructed angles with those known from the simulation.
Figures~\ref{fig:i_fig9} and \ref{fig:j_fig10} illustrate the effect of the two procedures.
In our case, the chosen values were, respectively, 55\% for the fraction of the length kept to determine its direction, and 1\% for the threshold above which ``beam'' voxels were also considered ``ejectile'' voxels.
Restricting only to a fraction of the ejectile track (\(\beta\) selection) mostly affects the precision of the reconstruction of the scattering angle: in Fig.~\ref{fig:i_fig9}b, the standard deviation of \( \theta_{\text{labs}} - \theta_{\text{labr}} \)\footnote{$\theta_{\text{labs}}$ = simulated laboratory angle, $\theta_{\text{labr}}$ = reconstructed laboratory angle} illustrates the improvement. The improved precision at higher $\theta_\text{c.m.}$ is due to the increasing length of the ejectile tracks. Including the voxels in the attenuation zone, on the other hand, mostly improves accuracy, as shown in Fig.~\ref{fig:j_fig10}c.


\section{Evaluation and Comparison}
\label{evalandcompare}

As a baseline for evaluation and comparison, we use the RANSAC method introduced above, because it is a reference for track reconstruction in active targets.
We experimented with the RANSAC method using distance threshold values between 5~mm and 10~mm, as well as with variants in which points selected in one model could be reused in subsequent models.
In the latter case, different suppression factors ranging from 0 to 1 were tested.
From these studies, we selected the parameter set and variant that provided the best performance for $\theta_\mathrm{c.m.} = 1^{\circ}$, corresponding to small-angle, short-range tracks.
This choice is motivated by the need to specifically observe the differences with the GMM when, for the latter, voxels inside the attenuation zone are considered for the ejectile tracks.

The parameters eventually used for the RANSAC analysis were: 5~mm for the distance threshold, 5000 for the number of iterations, and $s=10$ as the minimum number of voxels in each track.
The latter was then imposed on the GMM analysis as well.

We evaluated and compared the performance of RANSAC and GMM by using two complementary approaches.
The first was based on the \emph{quality of the clustering}, for which we used the Adjusted Rand Index (ARI) \cite{hubert1985comparing} to quantify the agreement between the reconstructed clusters and the simulated ground-truth track assignments.
ARI provides a measure of clustering consistency, independent of the physics observable, and is sensitive to both over-splitting and over-merging of tracks.
The second approach considered the accuracy of the \emph{reconstruction of the physical event} by comparing the reconstructed laboratory angle $\theta_{\mathrm{labr}}$ with the simulated quantity $\theta_{\mathrm{labs}}$.

\subsection{Evaluation of clustering performance using the ARI score}
\label{evaluate_ari}

A common way to evaluate the performance of the applied clustering technique is to consider pairs of points and measure their similarity with respect to the true clustering structure.
The clusters known from the input of the simulation were labeled as $t$ (``true'' clusters, with $t=1,\ldots,T$), while those reconstructed with the GMM clustering technique were labeled as $p$ (``predicted'' clusters, $p=1,\ldots,P$).
While the Rand index \cite{Rand71} is a convenient measure of similarity, it does not account for the random splits that might occur during clustering.
The ARI (Adjusted Rand Index) \cite{hubert1985comparing} score, defined as 
\begin{equation}
	\mathrm{ARI} = \frac{\sum_{t,p} {\binom{n_{tp}}{2}} - \left[\sum_t {\binom{n_t}{2}} \sum_p {\binom{n_p}{2}}\right]/\binom{n}{2}}{\frac{1}{2}\left[\sum_t {\binom{n_t}{2}} + \sum_p {\binom{n_p}{2}}\right] - \left[\sum_t {\binom{n_t}{2}} \sum_p {\binom{n_p}{2}}\right]/\binom{n}{2}}
\end{equation}
penalizes random splits and hence measures the overall performance of the technique in predicting the true structure.
$n_{tp}$ represents the number of elements that occur both in the true cluster $t$ and the predicted cluster $p$; thus, 
$\binom{n_{tp}}{2}$ is the number of pairs that can be formed by those elements (with the definition $\binom{n_{tp}}{2}=0$ when $n_{tp}=0$ or $1$).
$n_t$ and $n_p$ represent the number of elements in clusters $t$ and $p$, respectively, with $\binom{n_t}{2}$ and $\binom{n_p}{2}$ the respective number of pairs; finally, $\binom{n}{2}$ is the total number of pairs of elements.

\begin{figure}[t]
	\centering
	\medskip
	\includegraphics[width=1.0\textwidth]{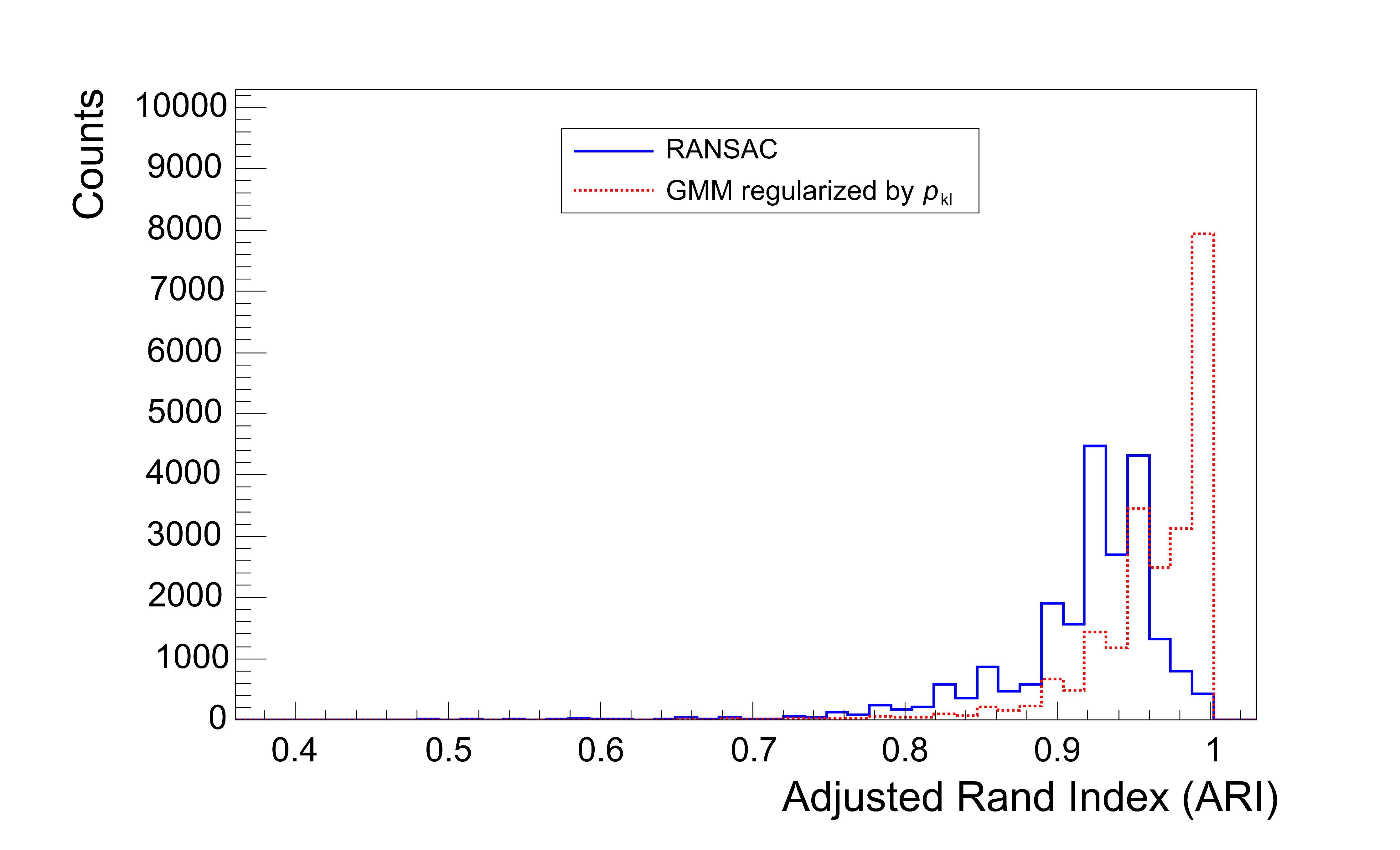}
	\caption[Short caption for Table of Figures]{
	Histogram showing the Adjusted Rand Index scores (1 = perfect agreement) for the simulated dataset.}
	\label{fig:k_fig11}
\end{figure}

Fig.~\ref{fig:k_fig11} shows the recorded ARI scores for the simulated dataset.
The histogram shown in red represents the ARI score distribution for all events obtained after regularization of the clusters identified by the GMM algorithm using the metric \( p_{kl} \).
The blue histogram corresponds to the ARI scores obtained using RANSAC.

We observe a limited improvement (not shown) in ARI index after regularization by $c_{kl}$.
In our data, the beam cluster contains many more points than the ejectile, so the majority of point pairs are beam–beam or beam–ejectile pairs.
These dominant pairs overwhelm the contribution of ejectile–ejectile pairs in the ARI computation, making the index largely insensitive to improvements in the small ejectile cluster, even when fragmented tracks are successfully merged.
The slightly left-shifted histogram of ARI indices for RANSAC-derived events indicates that the GMM achieves more consistent and correct assignment of points near the beam–ejectile interface, where RANSAC tends to misclassify a few boundary points.
The results also demonstrate that GMM, after regularization, can achieve nearly 100\% correct assignment of the voxels to the original simulated clusters.

\subsection{Comparison of track reconstruction performance}
\label{compare_ransac}

The comparison between the track reconstruction by GMM and RANSAC was made on a subset of all simulated data.
In particular, we considered only events for which the reconstruction found one ejectile track with at least 10 voxels outside the attenuation zone, and for which the reaction vertex and the start and end points of the ejectile track lay within the active volume of the detector or at most 10~mm upstream\footnote{In ACTAR TPC the full gas volume extends a few centimeters beyond the active volume.}.
Finally, to define an efficiency at each center-of-mass angle, we only retained events for which the reconstructed scattering angle $\theta_{\mathrm{labr}}$ was within 20 degrees of the correct (simulated) one $\theta_{\mathrm{labs}}$.

\begin{figure}
	\centering
	\medskip
	\includegraphics[width=1.0\textwidth]{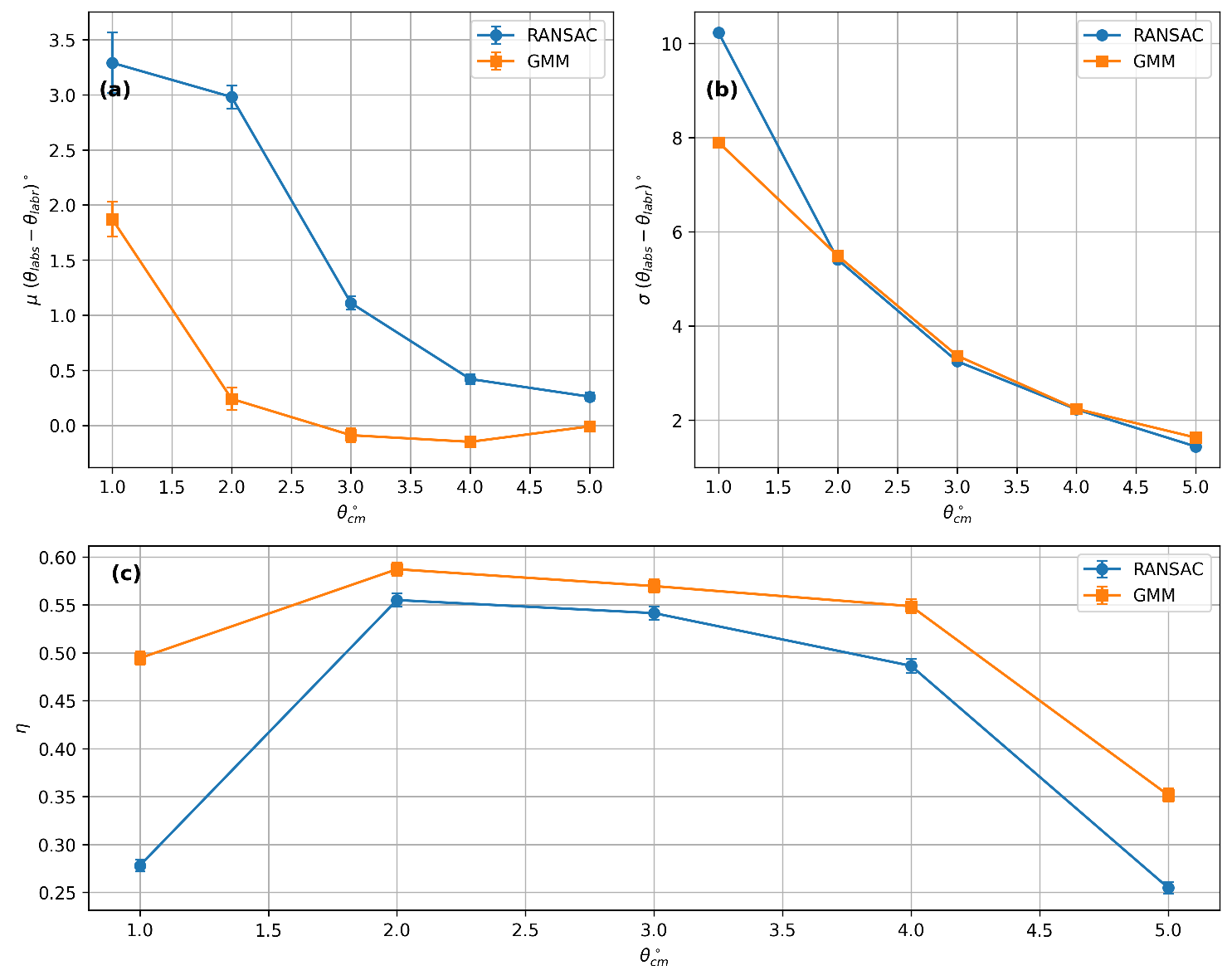}
	\caption[Short caption for Table of Figures]{Comparison of track reconstruction performance between GMM (regularized by $p_{kl}$, $c_{kl}$) and RANSAC (regularized by $c_{kl}$):
    (a) mean of the difference $\mu = \langle \theta_{\rm labs} - \theta_{\rm labr} \rangle$ as function of the simulated center-of-mass angle $\theta_\mathrm{c.m.}$; error bars represent the standard error of this mean, $\sigma/\sqrt{M_\mathrm{sel}}$, where $\sigma$ is the standard deviation of the selected events and $M_\mathrm{sel}$ is the number of selected events;
    (b) standard deviation of this difference;
    (c) reconstruction efficiency $\eta$, defined as the fraction of simulated tracks satisfying the selection criteria described in the text; error bars represent the binomial standard deviation, $\sqrt{\eta(1-\eta)/M}$, where $M$ is the total number of simulated events.} 
	\label{fig:l_fig12}
\end{figure}

Figure~\ref{fig:l_fig12} summarizes the comparison between GMM (regularized by the $p_{kl}$ and $c_{kl}$ metrics) and RANSAC (regularized by the $c_{kl}$).

For small-angle tracks ($\theta_\mathrm{c.m.} = 1^\circ$), GMM outperforms RANSAC in accuracy (Fig.~\ref{fig:l_fig12}a), precision (panel b), and efficiency\footnote{The absolute efficiency has little significance here because it refers to the full simulation data set.} (c), reflecting its ability to robustly reconstruct short, small-angle tracks.
This performance is enhanced by the possibility, in GMM, of incorporating points inside the attenuation zone based on the posterior probability, 
since the track direction is primarily determined by the initial points of the ejectile track.
An analysis of the events correctly reconstructed by the GMM but not detected by RANSAC reveals that the primary discriminating factor in these cases is the absolute difference between the simulated and reconstructed angles.

For larger-angle tracks (for example $\theta_\mathrm{c.m.} = 5^\circ$), the GMM still demonstrates superior overall performance, primarily because RANSAC tends to oversplit tracks at the 5~mm threshold.
Increasing the RANSAC threshold to 7~mm slightly improves its efficiency, resulting in nearly-equal performance for center-of-mass angles of $\theta_\mathrm{c.m.} = 2^\circ$, $3^\circ$, and $4^\circ$, and a marginally higher efficiency for RANSAC at $\theta_\mathrm{c.m.} = 5^\circ$.
However, this adjustment significantly degrades the small-angle reconstruction performance of RANSAC, with severe efficiency losses observed at $\theta_\mathrm{c.m.} = 1^\circ$.

Overall, the DBSCAN+GMM+merging approach provides a balanced reconstruction method, particularly in scenarios where the detection of short-range, small-angle tracks is critical.

\subsection{Computational Comparison}
To assess the computational cost of the DBSCAN + GMM + merging workflow for our simulated dataset, we record the run-time of each step and compare the total runtime with that of the RANSAC baseline. The two parameters that primarily determine the GMM runtime are the maximum number of EM (Expectation-Maximization) iterations and the tolerance used for convergence. In our case, we use the default values provided in the \textsc{scikit-learn} implementation of the Gaussian Mixture Model \cite{scikit-learn}. We observe that the default maximum number of iterations (100) is never reached; instead, the stopping criterion based on default tolerance ($10^{-3}$) is almost always satisfied earlier. This indicates that the tolerance parameter is the dominant factor governing the computational cost of GMM in our application. The computational time also depends on the total number of voxels $N_\text{tot}$ in each event and the number of clusters $K$. The combined median run-times ($\tilde{t}$ = median($t_m$), $t_m$ being the runtime per event, $m = 1,\ldots,M$ where M is the total number of events), measured using Python's \texttt{perf\_counter()}, for DBSCAN, GMM, and the $p_{kl}$-based merging are reported in Table~\ref{tab:timing_summary}. The median run-times for DBSCAN and $p_{kl}$-based merging are approximately 10~ms and the dominant computational cost in our workflow comes from GMM.

\begin{table}[tbp]
\begin{center}
\caption{Median runtimes ($\tilde{t}$) for RANSAC (for 1000 and 5000 iterations) measured in seconds, comapred to the  combined median run times for DBSCAN, GMM and merging based on $p_{kl}$, for center-of-mass angles ($\theta_\text{c.m.}$) ranging from 1$^\circ$ to 5$^\circ$. The number of voxels ($N_\text{tot}$) per event increases with increasing $\theta_\text{c.m.}$ due to the increasing track length of the ejectile tracks. Errors correspond to Median Absolute Deviation.}
\label{tab:timing_summary}
\begin{tabular}{cccc}
\hline\hline
$\theta_{\rm cm}$ (deg) & RANSAC$_{1000}$ (s) & RANSAC$_{5000}$ (s) & DBSCAN+GMM+merge (s) \\
\hline
1 & 0.60 $\pm$ 0.01 & 2.95 $\pm$ 0.04 & 1.19 $\pm$ 0.10 \\
2 & 0.60 $\pm$ 0.01 & 2.99 $\pm$ 0.03 & 1.17 $\pm$ 0.13 \\
3 & 0.57 $\pm$ 0.01 & 2.83 $\pm$ 0.04 & 1.21 $\pm$ 0.20 \\
4 & 0.61 $\pm$ 0.02 & 3.00 $\pm$ 0.09 & 1.32 $\pm$ 0.25 \\
5 & 0.62 $\pm$ 0.09 & 2.96 $\pm$ 0.21 & 1.51 $\pm$ 0.32 \\
\hline\hline
\end{tabular}
\end{center}
\end{table}

For RANSAC, the runtime depends on $N_\text{tot}$, $K$, the minimum number of voxels $s$, and the number of iterations $N_\text{iter}$. 
The latter is based on the standard estimate \cite{fischler1981random}:
\begin{equation}
\label{eq:ransac_iterations}
N_\text{iter}  = \frac{\ln(1-z)}{\ln(1-b)}
\end{equation}
where $z$ is the desired probability of selecting at least one set of voxels belonging to a single true track, and $b=w^s$, with $w$ the probability of the voxel belonging to a selected track and $s$ the minimum number of voxels. Based on this formula, we estimate the required number of iterations to be between 1000 and 2000. In cases where track fragmentation reduces the fraction of voxels belonging to the ejectile track with respect to the remaining  number of voxels in the event (assuming minimal noise in the simulated data), the number of iterations required can increase up to 5000. These are heuristic estimates.

For the efficiency calculations, we set the maximum number of RANSAC iterations to 5000, prioritizing accuracy. We report median runtimes for both 1000 and 5000 iterations, with uncertainties given by the Median Absolute Deviation (MAD)\footnote{$
\mathrm{MAD} = \mathrm{median}\left( t_m - \tilde{t} \right), \ \ m = 1, \dots, M$.}. Overall, we observe that the combined approach---DBSCAN followed by GMM clustering and the $p_{kl}$-based merging---achieves runtimes of the same order as RANSAC.


\section{Conclusion}
\label{conclusion}

We have presented a novel approach for track reconstruction in active targets that combines DBSCAN clustering with Gaussian Mixture Model (GMM) fitting and merging based on the $p$-value of the Mahalanobis distance between clusters.
This method proves particularly effective for short, small-angle tracks, which are notoriously difficult to reconstruct using conventional sequential RANSAC due to partial overlap with the beam tracks and deviations from rectilinear shape due to low-energy scattering. 
Compared to RANSAC, the GMM-based approach yields improved accuracy and efficiency for small-angle tracks, notably at $\theta_{\rm c.m.} = 1^\circ$, where it achieves better performance in both the mean and standard deviation of reconstructed laboratory angles, as well as in reconstruction efficiency.
A key advantage lies in its ability to include points within the attenuation zone through posterior-probability weighting, which enhances track completeness.
The method also demonstrates robust handling of low-energy track deviations, incorporating scattered points and non-rectilinear trajectories in a principled manner.
At higher $\theta_\mathrm{c.m.}$ values, GMM fitting maintains competitive performance, whereas RANSAC tends to oversplit tracks when lower thresholds are applied to preserve short-track sensitivity.

Overall, the combination of DBSCAN clustering, GMM fitting, and Mahalanobis $p$-value merging provides a balanced and automated reconstruction strategy capable of treating scattered points and short tracks consistently within a probabilistic framework.

In summary, this work shows that probabilistic clustering techniques such as GMM can significantly improve track reconstruction in low-energy, short-range regimes, leading to more accurate angular and efficiency measurements in active-target detectors.
Future studies will focus on extending this approach to higher-multiplicity events and real experimental datasets.

\section{Acknowledgments}
\label{acknowledgments}

The research leading to these results has received funding from the European Union's Horizon 2020 research and innovation programme under grant agreement no. 654002; from the Research Foundation – Flanders under research project G0A0419N and projects I002219N and I001323N of the International Research Infrastructure programme; from the KU Leuven internal funds under project C14/22/104; and from P2IO LabEx (ANR-10-LABX-0038) in the framework ``Investissements d'Avenir'' (ANR-11-IDEX-0003-01) managed by the Agence Nationale de la Recherche (ANR), France.

\bibliographystyle{elsarticle-num} 
\bibliography{references}

@inbook{Bishop2006_Ch9,
  title    = {Pattern {R}ecognition and {M}achine {L}earning},
  author       = {Christopher M. Bishop},
  year         = 2006,
  publisher    = {Springer {N}ew {Y}ork, {NY}},
  chapter      = {9},
  pages        = {423--455},
  edition      = {First}
}

@article{mauss2019commissioning,
Author = {Mauss, B. and Morfouace, P. and Roger, T. and Pancin, J. and Grinyer, G.
   F. and Giovinazzo, J. and Alcindor, V and Alvarez-Pol, H. and Arokiaraj,
   A. and Babo, M. and Bastin, B. and Borcea, C. and Caamano, M. and
   Ceruti, S. and Fernandez-Dominguez, B. and Foulon-Moret, E. and
   Gangnant, P. and Giraud, S. and Laffoley, A. and Mantovani, G. and
   Marchi, T. and Monteagudo, B. and Pibernat, J. and Poleshchuk, O. and
   Raabe, R. and Refsgaard, J. and Revel, A. and Saillant, F. and Stanoiu,
   M. and Wittwer, G. and Yang, J.},
Title = {Commissioning of the ACtive TARget and Time Projection Chamber ({ACTAR
   TPC)}},
Journal = {Nucl. Instrum. Methods Phys. Res. A},
Year = {2019},
Volume = {940},
Pages = {498-504},
DOI = {10.1016/j.nima.2019.06.067},
}

@article{blaizot1980nuclear,
  title={Nuclear compressibilities},
  author={Blaizot, Jean-Paul},
  journal={Physics Reports},
  volume={64},
  number={4},
  pages={171--248},
  year={1980},
  publisher={Elsevier},
  DOI = {10.1016/0370-1573(80)90001-0},
}

@article{garg2018compression,
	title={The compression-mode giant resonances and nuclear incompressibility},
	author={Garg, Umesh and Colo, Gianluca},
	journal={Progress in Particle and Nuclear Physics},
	volume={101},
	pages={55--95},
	year={2018},
	publisher={Elsevier},
  DOI = {10.1016/j.ppnp.2018.03.001},
}

@article{khan2011low,
	title={Low-energy monopole strength in exotic nickel isotopes},
	author={Khan, E and Paar, Nils and Vretenar, Dario},
	journal={Physical Review C},
	volume={84},
	number={5},
	pages={051301},
	year={2011},
	publisher={APS},
    DOI = {10.1103/PhysRevC.84.051301},
}

@article{gambacurta2019soft,
  title={Soft breathing modes in neutron-rich nuclei with the subtracted second random-phase approximation},
  author={Gambacurta, D and Grasso, M and Sorlin, O},
  journal={Physical Review C},
  volume={100},
  number={1},
  pages={014317},
  year={2019},
  publisher={APS},
  DOI = {10.1103/PhysRevC.100.014317},
}

@ARTICLE{Mahalanobis18,
title= "{Reprint of: Mahalanobis, P.C. (1936) ``On the Generalised Distance in Statistics.''}",
journal="Sankhya A",
volume= "80 (Suppl 1)",
pages=  "1-7",
year=   "2018",
doi="10.1007/s13171-019-00164-5",
    }

@ARTICLE{Gio96,
author= "Y. Giomataris and Ph. Rebourgeard and J.P. Robert and G. Charpak",
title=  "MICROMEGAS: a high-granularity position-sensitive gaseous detector for high
        particle-flux environments",
Journal = {Nucl. Instrum. Methods Phys. Res., A},
volume= "376",
pages=  "29",
year=   "1996",
doi="10.1016/0168-9002(96)00175-1",
    }

@article{fischler1981random,
	title={Random sample consensus: a paradigm for model fitting with applications to image analysis and automated cartography},
	author={Fischler, Martin A and Bolles, Robert C},
	journal={Communications of the ACM},
	volume={24},
	number={6},
	pages={381--395},
	year={1981},
	publisher={ACM New York, NY, USA},
    doi={10.1145/358669.358692},
}

@article{ayyad2018novel,
	title={Novel particle tracking algorithm based on the random sample consensus model for the active target time projection chamber (AT-TPC)},
	author={Ayyad, Yassid and Mittig, Wolfgang and Bazin, Daniel and Beceiro-Novo, Saul and Cortesi, Marco},
	journal={Nuclear Instruments and Methods in Physics Research Section A: Accelerators, Spectrometers, Detectors and Associated Equipment},
	volume={880},
	pages={166--173},
	year={2018},
	publisher={Elsevier},
    doi={10.1016/j.nima.2017.10.090},
}

@article{dempster1977maximum,
	title={Maximum {L}ikelihood from {I}ncomplete {D}ata via the {EM} {A}lgorithm},
	author={Dempster, Arthur P and Laird, Nan M and Rubin, Donald B},
	journal={Journal of the royal statistical society: series B (methodological)},
	volume={39},
	number={1},
	pages={1--22},
	year={1977},
	publisher={Wiley Online Library},
    url={http://www.jstor.org/stable/2984875},
}

@article{scikit-learn,
	title={Scikit-learn: Machine Learning in {P}ython},
	author={Pedregosa, F. and Varoquaux, G. and Gramfort, A. and Michel, V.
	and Thirion, B. and Grisel, O. and Blondel, M. and Prettenhofer, P.
	and Weiss, R. and Dubourg, V. and Vanderplas, J. and Passos, A. and
	Cournapeau, D. and Brucher, M. and Perrot, M. and Duchesnay, E.},
	journal={Journal of Machine Learning Research},
	volume={12},
	pages={2825--2830},
	year={2011},
	url={https://dl.acm.org/doi/10.5555/1953048.2078195}
}

@inproceedings{Ester1996density,
author = {Ester, Martin and Kriegel, Hans-Peter and Sander, J\"{o}rg and Xu, Xiaowei},
title = {A density-based algorithm for discovering clusters in large spatial databases with noise},
year = {1996},
publisher = {AAAI Press},
booktitle = {Proceedings of the Second International Conference on Knowledge Discovery and Data Mining},
pages = {226–231},
numpages = {6}, 
location = {Portland, Oregon},
series = {KDD'96},
url = {https://dl.acm.org/doi/10.5555/3001460.3001507}
}

@article{hubert1985comparing,
	title={Comparing partitions},
	author={Hubert, Lawrence and Arabie, Phipps},
	journal={Journal of classification},
	volume={2},
	number={1},
	pages={193--218},
	year={1985},
	publisher={Springer}
}

@article{ziegler2010srim,
	title={SRIM--The stopping and range of ions in matter (2010)},
	author={Ziegler, James F and Ziegler, Matthias D and Biersack, Jochen P},
	journal={Nuclear Instruments and Methods in Physics Research Section B: Beam Interactions with Materials and Atoms},
	volume={268},
	number={11-12},
	pages={1818--1823},
	year={2010},
	publisher={Elsevier}
}

@article{delchambre2015weighted,
  title={Weighted principal component analysis: a weighted covariance eigendecomposition approach},
  author={Delchambre, Ludovic},
  journal={Monthly Notices of the Royal Astronomical Society},
  volume={446},
  number={4},
  pages={3545--3555},
  year={2015},
  publisher={Oxford University Press}
}

@article{Matta_2016,
	doi = {10.1088/0954-3899/43/4/045113},
	year = 2016,
	month = {mar},
	publisher = {{IOP} Publishing},
	volume = {43},
	number = {4},
	pages = {045113},
	author = {A Matta and P Morfouace and N de S{\'{e}}r{\'{e}}ville and F Flavigny and M Labiche and R Shearman},
	title = {{NPTool}: a simulation and analysis framework for low-energy nuclear physics experiments},
	journal = {Journal of Physics G: Nuclear and Particle Physics},
}

@book{hartigan1975clustering,
	title={Clustering algorithms},
	author={Hartigan, John A},
	year={1975},
	publisher={John Wiley \& Sons, Inc.}
}

@article{press1990savitzky,
	title={Savitzky-{G}olay smoothing filters},
	author={Press, William H and Teukolsky, Saul A},
	journal={Computers in Physics},
	volume={4},
	number={6},
	pages={669--672},
	year={1990},
	publisher={American Institute of Physics}
}

@book{harakeh2001giant,
  title={Giant Resonances: fundamental high-frequency modes of nuclear excitation},
  author={Harakeh, Muhsin N and Woude, Adriaan},
  volume={24},
  year={2001},
  publisher={Oxford Studies in Nuclear Phys}
}

@article{Rand71,
author = {William M. Rand},
title = {Objective Criteria for the Evaluation of Clustering Methods},
journal = {Journal of the American Statistical Association},
volume = {66},
number = {336},
pages = {846--850},
year = {1971},
publisher = {ASA Website},
doi = {10.1080/01621459.1971.10482356},
}

@article{Schwarz1978,
author = {Gideon Schwarz},
title = {{Estimating the Dimension of a Model}},
volume = {6},
journal = {The Annals of Statistics},
number = {2},
publisher = {Institute of Mathematical Statistics},
pages = {461 -- 464},
year = {1978},
doi = {10.1214/aos/1176344136},
}

@book{Fruhwith2000,
  author = {Fr{\"u}hwirth, R. and Regler, M. and Bock, R. K. and Grote, H. and Notz, D.},
  year = {2000},
  title = {Data Analysis Techniques for High-Energy Physics},
  edition = {2nd},
  publisher = {Cambridge University Press},
  address = {Cambridge},
  series = {Cambridge Monographs on Particle Physics, Nuclear Physics, and Cosmology},
  isbn = {0521632196}
}

@article{Bradt2017,
title = {Commissioning of the Active-Target Time Projection Chamber},
journal = {Nuclear Instruments and Methods in Physics Research Section A: Accelerators, Spectrometers, Detectors and Associated Equipment},
volume = {875},
pages = {65-79},
year = {2017},
issn = {0168-9002},
doi = {doi.org/10.1016/j.nima.2017.09.013},
author = {J. Bradt and D. Bazin and F. Abu-Nimeh and T. Ahn and Y. Ayyad and S. Beceiro Novo and L. Carpenter and M. Cortesi and M.P. Kuchera and W.G. Lynch and W. Mittig and S. Rost and N. Watwood and J. Yurkon},
}

@article{Dalitz2017,
    title   = {{Iterative Hough Transform for Line Detection in 3D Point Clouds}},
    author  = {Dalitz, Christoph and Schramke, Tilman and Jeltsch, Manuel},
    journal = {{Image Processing On Line}},
    volume  = {7},
    pages   = {184--196},
    year    = {2017},
    doi    = {10.5201/ipol.2017.208}
}

@article{ZAMORA2021,
title = {Tracking algorithms for TPCs using consensus-based robust estimators},
journal = {Nuclear Instruments and Methods in Physics Research Section A: Accelerators, Spectrometers, Detectors and Associated Equipment},
volume = {988},
pages = {164899},
year = {2021},
issn = {0168-9002},
doi = {10.1016/j.nima.2020.164899},
author = {J.C. Zamora and G.F. Fortino},
}

@article{Ayyad2023,
author = {Ayyad, Yassid and Anthony, Adam K. and Bazin, Daniel and Chen, Jie and McCann, Gordon W. and Mittig, Wolfgang and Kay, Benjamin P. and Sharp, David K. and Zamora, Juan Carlos},
	doi = {10.1140/epja/s10050-023-01205-2},
	journal = {The European Physical Journal A},
	number = {12},
	pages = {294},
	title = {Kinematics reconstruction in solenoidal spectrometers operated in active target mode},
	volume = {59},
	year = {2023},
}

@article{Dalitz2019,
title = {Automatic trajectory recognition in Active Target Time Projection Chambers data by means of hierarchical clustering},
journal = {Computer Physics Communications},
volume = {235},
pages = {159-168},
year = {2019},
issn = {0010-4655},
doi = {10.1016/j.cpc.2018.09.010},
author = {Christoph Dalitz and Yassid Ayyad and Jens Wilberg and Lukas Aymans and Daniel Bazin and Wolfgang Mittig},
}

@article{Bazin2020,
title = {Low energy nuclear physics with active targets and time projection chambers},
journal = {Progress in Particle and Nuclear Physics},
volume = {114},
pages = {103790},
year = {2020},
doi = {10.1016/j.ppnp.2020.103790},
author = {D. Bazin and T. Ahn and Y. Ayyad and S. Beceiro-Novo and A.O. Macchiavelli and W. Mittig and J.S. Randhawa},
}

@article{schubert2017dbscan,
  title={{DBSCAN} {R}evisited, {R}evisited: {W}hy and {H}ow {Y}ou {S}hould ({S}till) use {DBSCAN}},
  author={Schubert, Erich and Sander, J{\"o}rg and Ester, Martin and Kriegel, Hans Peter and Xu, Xiaowei},
  journal={ACM Transactions on Database Systems (TODS)},
  volume={42},
  pages={19},
  year={2017},
  publisher={Acm New York, NY, USA},
  doi = {10.1145/3068335},
}

\end{document}